\documentclass[a4paper,fleqn]{cas-sc}
\usepackage{graphicx} 
\usepackage[authoryear,longnamesfirst]{natbib}
\usepackage{bm}
\usepackage{natbib}
\usepackage{caption}
\setcitestyle{numbers,square}
\def\tsc#1{\csdef{#1}{\textsc{\lowercase{#1}}\xspace}}
\tsc{WGM}
\tsc{QE}
\tsc{EP}
\tsc{PMS}
\tsc{BEC}
\tsc{DE}

\begin{document}
\let\WriteBookmarks\relax
\def\floatpagepagefraction{1}
\def\textpagefraction{.001}
\shorttitle{}
\shortauthors{J. Ye et~al.}

\title [mode = title]{Fractional-Order Subband p-Norm Adaptive Filter via Transformation Nearest Kronecker Product Decomposition for Active Noise Control}

\tnotetext[footnotes]{This work was supported by the National Natural Science Foundation
	of China (NSF) under Grant 62171388, Grant 61871461, and Grant
	61571374, the East China Jiaotong University Scientific Research Fund under Grant 2003425029. Email addresses: yjh\_zcl@163.com (J. Ye), hqzhao\_swjtu@126.com (H. Zhao), shaohuilv\_swjtu@126.com (S. Lv), zhouyang\_swjtu@126.com (Y. Zhou). \\
	$*$ Corresponding author: Haiquan Zhao.\\}
\author{ Jianhong Ye, Haiquan Zhao$^{*}$, Shaohui Lv, and Yang Zhou}
\affiliation{organization={Key Laboratory of Magnetic Suspension Technology and Maglev Vehicle, Ministry of Education, School of Electrical Engineering, Southwest Jiaotong University},
	city={Chengdu},
	postcode={610031},
	country={China}}

\begin{abstract}
The conventional normalized subband p-norm (NSPN) algorithm achieves robustness in $\alpha$-stable noise ($1<\alpha \leq 2$) by utilizing low-order error moments. However, its performance degrades significantly under three scenarios: (1) non-Gaussian inputs, (2) $\alpha$-stable noise with $0<\alpha \leq 1$, and (3) sparse system identification. To address these limitations, this paper proposes a fractional-order NSPN algorithm based on the nearest Kronecker product (NKP) decomposition and fractional-order stochastic gradient descent, termed NKP-FoNSPN. Theoretical bounds for the fractional-order parameter $\beta$ are also derived. Notably, when $\beta=1$, the NKP-FoNSPN reduces to a new NKP-NSPN algorithm, while its non-NKP decomposition variant becomes the fractional-order NSPN (FoNSPN) algorithm. Furthermore, a novel transformation-based NKP (TNKP) decomposition technique is designed, which exhibits lower computational complexity than conventional NKP for specific filter structures. The resulting TNKP-based FoNSPN (TNKP-FoNSPN) achieves lower steady-state misadjustment and multiplication cost compared with the NKP-FoNSPN algorithm. Additionally, complete computational complexity analyses are provided. For active noise control (ANC) scenarios, we develop filtered-x variants: NKP-FxFoNSPN and TNKP-FxFoNSPN. From the former, two additional variants are derived: NKP-FxNSPN and FxFoNSPN. Simulations using diverse noise sources (pink, helicopter, gunshot, pile driver, and traction substation noise) demonstrate the superiority of the proposed algorithms. Finally, we validate their noise reduction performance in a real constructed single-channel duct ANC and a simulated multi-channel ANC systems.
	
\end{abstract}


\maketitle

\section{Introduction}
In recent decades, adaptive filtering technology has played an irreplaceable role in key areas such as active noise control (ANC) \cite{wang2025family, shen2025multi, li2024enhanced,li2026transform}, radar signal processing \cite{spafford2003optimum}, integrated systems \cite{mula2017algorithm}, blind source separation \cite{10510496}, system identification \cite{zhao2013identification, zhao2011low, zhao2014memory, zhao2022robust,zhao2022robust1,peng2025family,zhao2008functional,zhao2010adaptive,zhao2009adaptively,ye2025nearest,ye2025p,peng2026fast}, and artificial intelligence \cite{feng2025meta, 10323505}. Among numerous adaptive filtering algorithms, the normalized least mean square (NLMS) algorithm \cite{236504} is widely adopted in these scenarios owing to the smoothness, convexity, and optimality of the mean square error (MSE) criterion under Gaussian assumptions. However, the convergence rate of the NLMS algorithm deteriorates significantly when handling highly correlated input signals. To overcome this limitation, researchers developed the normalized subband adaptive filtering (NSAF) algorithm \cite{1324714}, which decomposes highly correlated input signals into multiple subband input signals via analysis filter banks. This decomposition enables subband input signals to exhibit near-white statistical properties, significantly improving convergence performance while maintaining computational complexity comparable to that of the NLMS algorithm due to the critically sampling strategy. Notably, the stability of both the NLMS and NSAF algorithms is highly dependent on Gaussian noise assumptions. Nevertheless, non-Gaussian noise (e.g., impulsive noise) present in real-world applications - such as SI \cite{10806852}, acoustic echo cancellation (AEC) \cite{8098540}, and speech enhancement \cite{9468934} - poses severe challenges to MSE-based adaptive filters, as its statistical properties deviate significantly from Gaussian models. Such noise is typically modeled using symmetric $\alpha$-stable random processes, with characteristic functions expressed as \cite{231338}
\begin{align}
	\label{000} 
	\phi(t)=\text{exp}\big(-\gamma \lvert t\lvert^{\alpha}\big).
\end{align}
Here, $\gamma>0$ is the scale parameter, and the characteristic exponent $\alpha \in(0,2]$ controls the impulsiveness of the noise. Cases $\alpha=1$ and $\alpha=2$ correspond to Cauchy and Gaussian noise, respectively. 

Numerous robust approaches have been proposed to tackle the performance degradation resulting from impulsive noise. For instance, the M-estimate NSAF (MNSAF) \cite{8888205} algorithm suspends weight coefficient updates during error outliers to maintain stable convergence. However, its performance degrades or diverges when continuous outliers occur. Moreover, information-theoretic learning (ITL)-based adaptive filters provide an alternative strategy for handling non-Gaussian noise, including the maximum correntropy criterion (MCC) \cite{5178823}, generalized MCC (GMCC) \cite{11111}, and minimum error entropy (MEE) \cite{8937723}. These algorithms capture the higher-order statistics of signals rather than merely their energy, rendering them effective against various impulsive noise types. To further improve the performance of GMCC, the kernel risk-sensitive loss (KRSL) has been proposed in \cite{11112}. To reduce the computational overhead of ITL-based adaptive filtering algorithms, the maximum versoria criterion (MVC) has been introduced in \cite{7858738}. Alternatively, the sign algorithms that exploit low-order error moments \cite{8863930, LIU2023103981} achieve stable convergence in non-Gaussian noise environments but neglect the amplitude information of the error signals, resulting in poor convergence. To utilize both magnitude and phase information of the error signals, researchers developed the normalized least mean $p$-norm (NLMP) \cite{335063} and normalized subband $p$-norm (NSPN) \cite{11113} algorithms based on the mean $p$-power error (MPE) criterion. These algorithms exhibit a significantly improved convergence rate compared to conventional sign algorithms. Unfortunately, their performance will deteriorate sharply when subjected to $\alpha$-stable noise with $0<\alpha\leq 1$. Notably, the aforementioned robustness strategies are designed under the assumption that distorted noise is a non-Gaussian process, while the input is assumed to be a Gaussian process. Consequently, the performance of these robust algorithms deteriorates when encountering non-Gaussian inputs. 

To tackle non-Gaussian inputs and $\alpha$-stable noise with $0<\alpha\leq 1$, fractional-order calculus offers a promising framework \cite{jumarie2013}. Adaptive filters based on fractional-order stochastic gradient descent (FoSGD) have demonstrated favorable performance in such noisy scenarios \cite{1,2,3,4,5,6,7,8,9}. In \cite{3}, the authors combined the FoSGD method with the NLMP algorithm to develop the fractional-order NLMP (FoNLMP) algorithm, which resolves the performance degradation issue of the NLMP algorithm in scenarios with $0<\alpha\leq 1$ while maintaining robustness to non-Gaussian inputs. Additionally, researchers have developed the fractional-order M-estimate adaptive filtering (FoMAF) \cite{1} and fractional-order MCC (FoMCC) \cite{5} algorithms to achieve stable convergence under both non-Gaussian inputs and $\alpha$-stable noise with $0<\alpha\leq 1$. To further enhance the overall performance of the FoMCC algorithm, the fractional-order MVC (FoMVC) \cite{8} and fractional-order hyperbolic tangent (FoHT) \cite{4} algorithms have been proposed. To address the steady-state fluctuation problem of fractional-order filters, Cui et al. introduced the fractional-order generalized Cauchy kernel loss (FoGCKL) and enhanced batch FoGCKL (EB-FoGCKL) algorithms \cite{9}.

In many applications related to the echo environments, the duration of the system impulse response (IR) to be estimated (such as room impulse response (RIR) or underwater acoustic channel response) is usually very long \cite{kuttruff2024room}. To accurately model this response, the length of the adaptive FIR filter needs to be increased accordingly, which significantly increases the computational complexity of the algorithm. Meanwhile, a long IR is usually accompanied by an increase in the eigenvalue spread of the autocorrelation matrix of the input signal \cite{h2a}, resulting in a decrease in the convergence rate of the filter. Recently, the nearest Kronecker product (NKP) decomposition technique based on tensor decomposition theory can decompose long IRs into Kronecker products of multiple low-rank sub-components \cite{van2000ubiquitous}, thereby improving the convergence performance of adaptive FIR filters \cite{8369106}. Consequently, by combining the NKP decomposition technique with the conventional NLMS algorithm, the resulting NKP-based NLMS (NKP-NLMS) algorithm \cite{9053421} achieves a faster convergence rate than the NLMS algorithm in low-rank systems. In \cite{8682498}, the recursive least squares (RLS)-based NKP (RLS-NKP) algorithm achieves convergence performance comparable to that of the RLS algorithm with significantly reduced computational complexity. Furthermore, the superior performance of a class of robust algorithms based on RLS-NKP and the low-complexity dichotomous coordinate descent (DCD) algorithm built upon it has been fully validated in scenarios involving active impulsive noise control (AINC) \cite{wang2025family}. Additionally, Bhattacharjee et al. proposed the NKP-based generalized hyperbolic secant adaptive filtering (NKP-GHSAF) algorithm, which achieves a faster convergence rate than the NKP-based GMCC (NKP-GMCC) algorithm in impulsive noise environments \cite{9170797}. Overall, this popular NKP decomposition technique has been widely used in adaptive beamforming \cite{KUHN2021102968}, Kalman filtering \cite{DOGARIU2020}, nonlinear filtering \cite{9444125}, and speech dereverberation \cite{9739987}.

In ANC applications, adaptive ANC systems generate anti-noise with opposite phases and equal amplitudes to counteract the irritating noise, thereby achieving noise control. The filtered-x least mean square (FxLMS) algorithm is widely used due to its low computational complexity, low latency, and simple structure \cite{kuo1996active, elliott2000}. However, the FxLMS algorithm based on the MSE criterion cannot achieve effective noise reduction for impulsive noise. Consequently, researchers have designed robust adaptive ANC algorithms for AINC scenarios. For instance, Leahy et al. developed the filtered-x least mean $p$-norm (FxLMP) algorithm based on the $p$-order error moment \cite{479472}, which achieves better impulsive noise reduction performance than the FxLMS algorithm. In \cite{kurian2017robust}, the filtered-x MCC (FxMCC) algorithm based on ITL was proposed to address noise sources modeled as impulsive noise. Furthermore, the filtered-x GMCC (FxGMCC) algorithm proposed in \cite{zhu2020robust} outperforms FxMCC in noise reduction. Recently, Zhou et al. developed the filtered-x affine-projection-like exponential hyperbolic sine (FxAPLEHS) algorithm \cite{zhou2024combined}, which achieves better noise control than the FxGMCC algorithm; however, its performance depends on the projection order. Typically, a higher projection order ensures better performance but incurs higher computational complexity. Additionally, the NKP-based multichannel filtered-x affine projection (NKP-MFxAP) algorithm \cite{li2025nearest} improves the convergence performance of the conventional multichannel filtered-x affine projection (MFxAP) algorithm in multichannel ANC systems. 
  
However, no studies have investigated NKP-based subband adaptive filtering algorithms, whether in SI, ANC, or other application scenarios, particularly robust subband NKP adaptive filtering algorithms against impulsive noise. Furthermore, the conventional NSPN algorithm exhibits good robustness against distorted noise modeled by $\alpha$-stable processes with $1<\alpha\leq 2$, but its performance deteriorates sharply or diverges under non-Gaussian inputs or $\alpha$-stable noise with $0<\alpha\leq 1$. Meanwhile, when the NSPN algorithm is used to estimate the IR of a sparse unknown system, it exhibits poor convergence performance. Thus, this paper develops algorithms to address these deficiencies. Our key contributions are as follows:

1) Leveraging the NKP decomposition technique and FoSGD framework, we propose a novel NKP decomposition-based fractional-order NSPN (NKP-FoNSPN) algorithm, and analyze the theoretical range of fractional-order parameter $\beta$. Overall, this algorithm fills the research gap regarding NKP decomposition-based subband adaptive filtering algorithms. Compared with the NSPN algorithm, the NKP-FoNSPN exhibits superior performance under both non-Gaussian inputs and $\alpha$-stable noise with $0<\alpha\leq 1$ as well as with $1<\alpha\leq 2$.
\begin{table*}[htp]
	\renewcommand\arraystretch{1}
	\tabcolsep = 2.3cm
	\setlength{\abovecaptionskip}{0cm}
	\setlength{\belowcaptionskip}{0cm}
	\begin{center}
		\caption{Summary of mathematical symbols.}
		\begin{tabular}{c c } 
			\hline
			\rowcolor{gray!30}\footnotesize{Notations} &Description \\			
			\hline
			$\otimes$ &Kronecker product \\
			\rowcolor{gray!30}$\lvert\lvert.\lvert\lvert_2$ &Euclidean norm \\
			$\lvert\lvert.\lvert\lvert_F$ &Frobenius norm \\
			\rowcolor{gray!30}$\lvert\lvert\bm{x}\lvert\lvert_p^p$ &$\sum_{i=1}^D\lvert x_i\lvert^p$ \\
			$(.)^{\text T}$ &Transpose of a vector or matrix\\
			\rowcolor{gray!30}$\text{sgn}(.)$ &Sign function \\
			${\bm I}_{D\times 1}$ &Identity matrix with $D\times 1$ \\
			\rowcolor{gray!30}$\Gamma(.)$ &Gamma function \\
			$\text{Re}\{\bm{x}\}$ & Real part of a vector \\
			\rowcolor{gray!30}$\text{vec}(.)$ & Converts $D\times D$ matrix into an $D^2\times 1$ vector \\
			\hline
		\end{tabular}
	\end{center}
\end{table*}

2) When $\beta=1$, the proposed NKP-FoNSPN algorithm reduces to a novel NKP-based NSPN (NKP-NSPN) algorithm, while its non-NKP decomposition variant becomes the fractional-order NSPN (FoNSPN) algorithm. 

3) To reduce steady-state misadjustment in NKP-based adaptive filtering algorithms, we develop a novel transformation-based NKP (TNKP) decomposition technique. This technique is applied to the existing NKP-NLMS and newly proposed NKP-FoNSPN algorithms to obtain the TNKP-based NLMS (TNKP-NLMS) and TNKP-based FoNSPN (TNKP-FoNSPN) algorithms. Notably, both TNKP variants achieve lower steady-state misadjustment and computational complexity than their NKP counterparts.

4) We analyze the computational complexity of the proposed NKP-FoNSPN and TNKP-FoNSPN algorithms, as well as that of the comparison algorithms. Additionally, we perform the applications of the NKP-FoNSPN type algorithms in the SI and AEC.

5) In ANC applications, we introduce the filtered-x NKP-FoNSPN (NKP-FxFoNSPN) and filtered-x TNKP-FoNSPN (TNKP-FxFoNSPN) algorithms. From the former, we derive two additional variants: filtered-x NKP-NSPN (NKP-FxNSPN) and filtered-x FoNSPN (FxFoNSPN). We validate the noise reduction performance of all proposed algorithms using pink noise, helicopter noise, gunshot, and pile driver noise. Finally, we set up a real single-channel duct ANC and a simulated multi-channel ANC systems to demonstrate the effectiveness of proposed algorithms in controlling traction substation noise.
  
The paper is organized as follows. Section 2 briefly reviews the Kronecker product decomposition and subband SI model. Section 3 derives the NKP-FoNSPN and TNKP-FoNSPN algorithms, analyzes the theoretical range of fractional-order $\beta$ values, and gives the computational complexity of the different algorithms. Section 4 derives the NKP-FxFoNSPN and TNKP-FxFoNSPN algorithms in detail. Section 5 provides the simulation results of the proposed algorithms in SI, AEC, and ANC scenarios. Finally, Section 6 concludes the paper.

The notations used in the sequel are listed in Table 1.
\section{Brief Review}
\subsection{ Kronecker Product Decomposition}
\label{sec:guidelines}
Let $\bm{m}_0$ represent the real-valued IR of the unknown system to be identified, with dimensions $D=D_1\times D_2$ and $D_1> D_2$. We decompose $\bm{m}_0$ into $\bm{m}_0=[\bm{t}_1^{\text T}, \bm{t}_2^{\text T},..., \bm{t}_{D_2}^{\text T}]^{\text T}$, where each $\{\bm{t}_i\}_{i=1}^{D_2}$ is a short IR of size $D_1 \times 1$. These vectors $\{\bm{t}_i\}_{i=1}^{D_2}$ are assumed to be strongly linearly dependent with each other \cite{elisei2019recursive}. Following \cite{8369106}, $\bm{m}_0$ can be approximated by $\bm{m}_2\otimes \bm{m}_1=\text{vec}(\bm{m}_1\bm{m}_2^{\text T})$, where $\bm{m}_1$ and $\bm{m}_2$ are two sub-impulse responses of sizes $D_1\times1$ and $D_2\times1$, respectively. To evaluate the approximation accuracy between $\bm{m}_0$ and $\bm{m}_2\otimes \bm{m}_1$, we define a normalized misalignment metric as: 
\vspace{-0.2cm} 
\begin{equation}
	\begin{split}
		\begin{array}{rcl}
			\begin{aligned}
				\label{001}
				\  \omega\overset{\bigtriangleup}{=}\lvert\lvert\bm{m}_0-\bm{m}_2\otimes \bm{m}_1\lvert\lvert_2/\lvert\lvert\bm{m}_0\lvert\lvert_2.
			\end{aligned}
		\end{array}
	\end{split}
\end{equation}

Next, $\bm{m}_0$ can be reorganized in matrix form as $\bm{M}_0=[\bm{t}_1, \bm{t}_2,..., \bm{t}_{D_2}]$ with a size of $D_1\times D_2$. Therefore, the NKP approximation in \eqref{001} can be expressed as:
\begin{equation}
	\begin{split}
		\begin{array}{rcl}
			\begin{aligned}
				\label{002}
				\  \omega\overset{\bigtriangleup}{=}\lvert\lvert\bm{M}_0-\bm{m}_1\bm{m}_2^{\text T}\lvert\lvert_F/\lvert\lvert\bm{M}_0\lvert\lvert_F,
			\end{aligned}
		\end{array} 
	\end{split}
\end{equation}
In this formulation, our objective is to find appropriate $\bm{m}_1$ and $\bm{m}_2$ to minimize the normalized misalignment metric $\omega$. Interestingly, minimizing the Frobenius norm of $\omega(\bm{m}_1,\bm{m}_2)$ amounts to identifying the nearest rank-1 matrix to $\bm{M}_0$. Singular value decomposition (SVD) provides the mathematical foundation for this optimal approximation \cite{golub2013matrix}. Therefore, we have 
\begin{equation}
	\begin{split}
		\begin{array}{rcl}
			\begin{aligned}
				\label{003}
				\  \bm{M}_0=\bm{ H}_1\bm{\Sigma}\bm{H}_2^{\text T}=\sum_{i=1}^{D_2}\rho_i\bm{h}_{1,i}\bm{h}_{2,i}^{\text T}.
			\end{aligned}
		\end{array}
	\end{split}
\end{equation}
Here, $\bm \Sigma$ represents a $D_1\times D_2$ rectangular diagonal matrix containing the singular values of $\bm{M}_0$ sorted in descending order ($\rho_1\geq\rho_2\geq...\geq\rho_{D_2}\geq0$). The matrices $\bm{ H}_1  \in \mathbb{R}^{D_1 \times D_1} $ and $\bm{ H}_2 \in \mathbb{R}^{D_2 \times D_2}$ contain the left and right singular vectors of $\bm{M}_0$, respectively, with $\bm{h}_{1,i}$ and $\bm{h}_{2,i}$ denoting their $i$-th columns. With them, the optimal IRs of ${\bm m}_1$ and ${\bm m}_2$ are $\breve{\bm m}_1=\sqrt{\rho_1}\bm{h}_{1,1}$ and $\breve{\bm m}_2=\sqrt{\rho_1}\bm{h}_{2,1}$, respectively. However, when the segments $\{\bm{t}_i\}_{i=1}^{D_2}$ lack strong linear dependence on each other, we extend to a closest rank-$Q$ approximation ($Q \leq D_2$):
\begin{equation}
	\begin{split}
		\begin{array}{rcl}
			\begin{aligned}
				\label{004}
				\  \bm{m}_0&\approx\sum_{q=1}^Q\bm{m}_{2,q}\otimes\bm{m}_{1,q}=\text{vec}\Bigg(\sum_{q=1}^Q\bm{m}_{1,q}\bm{m}_{2,q}^{\text T}\Bigg)=\text{vec}\big(\bm{M}_1\bm{M}_2^{\text T}\big),
			\end{aligned}
		\end{array}
	\end{split}
\end{equation}
where $\bm{m}_{1,q} \in \mathbb{R}^{D_1\times 1} $ and $\bm{m}_{2,q} \in \mathbb{R}^{D_2 \times 1}$ represent weight vectors for the $q$-th component, and the matrices $\bm{M}_1 \in \mathbb{R}^{D_1\times Q}$ and $\bm{M}_2 \in \mathbb{R}^{D_2\times Q}$ are constructed as:
\begin{equation}
	\begin{split}
		\begin{array}{rcl}
			\begin{aligned}
				\label{005}
				\  {\bm M}_1 = \big[{\bm m}_{1,1}\;{\bm m}_{1,2}\;...\;  {\bm m}_{1,Q}\big],
			\end{aligned}
		\end{array}
	\end{split}
\end{equation}
\vspace{-0.8cm}
\begin{equation}
	\begin{split}
		\begin{array}{rcl}
			\begin{aligned}
				\label{006}
				\   {\bm M}_2 = \big[{\bm m}_{2,1}\;{\bm m}_{2,2}\;...\; {\bm m}_{2,Q}\big].
			\end{aligned}
		\end{array}
	\end{split}
\end{equation}

Therefore, the approximation problem in \eqref{002} can be equivalently transformed into minimizing the Frobenius norm error:
\begin{equation}
	\begin{split}
		\begin{array}{rcl}
			\begin{aligned}
				\label{007}
				\  \omega\overset{\bigtriangleup}{=}\lvert\lvert\bm{M}_0-\bm{M}_1 \bm{M}_2^{\text T}\lvert\lvert_F/\lvert\lvert\bm{M}_0\lvert\lvert_F,
			\end{aligned}
		\end{array}
	\end{split}
\end{equation}
which leads to the optimal solutions:
\begin{equation}
	\begin{split}
		\begin{array}{rcl}
			\begin{aligned}
				\label{008}
				\  \breve{\bm M}_1 =& \big[\breve{\bm m}_{1,1}\;\breve{\bm m}_{1,2}\;...\;  \breve{\bm m}_{1,Q}\big]=\big[\sqrt{\rho_1}\bm{h}_{1,1}, \sqrt{\rho_2}\bm{h}_{1,2},...,\sqrt{\rho_Q}\bm{h}_{1,Q}\big],
			\end{aligned}
		\end{array}
	\end{split}
\end{equation}
\vspace{-0.8cm}
\begin{equation}
	\begin{split}
		\begin{array}{rcl}
			\begin{aligned}
				\label{009}
				\  \breve{\bm M}_2 =& \big[\breve{\bm m}_{2,1}\;\breve{\bm m}_{2,2}\;...\;  \breve{\bm m}_{2,Q}\big]=\big[\sqrt{\rho_1}\bm{h}_{2,1}, \sqrt{\rho_2}\bm{h}_{2,2},...,\sqrt{\rho_Q}\bm{h}_{2,Q}\big].
			\end{aligned}
		\end{array}
	\end{split}
\end{equation}

Based on \eqref{008} and \eqref{009}, the optimal approximation of $\bm{m}_0$ is
\begin{equation}
	\begin{split}
		\begin{array}{rcl}
			\begin{aligned}
				\label{010}
				\  \breve{\bm m}_0 =\sum_{q=1}^Q\breve{\bm m}_{2,q}\otimes\breve{\bm m}_{1,q} =\sum_{q=1}^Q\rho_q{\bm h}_{2,q}\otimes{\bm h}_{1,q}.
			\end{aligned}
		\end{array}
	\end{split}
\end{equation}
\subsection{Subband System Identification Model}
Within the SI framework, the input-output relationship of the unknown system  $\bm{m}_0$ at time instant $k$ is described by:
\begin{equation}
	\begin{split}
		\begin{array}{rcl}
			\begin{aligned}
				\label{011}
				\  \ d_k=\bm{x}^{\text T}_k\bm{m}_0+v_k,
			\end{aligned}
		\end{array}
	\end{split}
\end{equation}
where $d_k$ denotes the desired signal, $\bm{x}_k=[x_k,x_{k-1},...,x_{k-D+1}]^{\text T} \in \mathbb{R}^{D \times 1}$ represents the input vector, and $v_k$ is the additive noise.

In the subband structure, the fullband signals $\bm{x}_k$ and $d_k$ are filtered through analysis filters with impulse responses $\{\bm{f}_j\}_{j=1}^N$ with $L$-taps ($N$ subbands), yielding the subband input vector $\bm{x}_{k,j}=[x_{k,j},x_{k-1,j},...,x_{k-D+1,j}]^{\text T}$ and the subband desired signal $d_{k,j}$, respectively. The subband error signal $e_{k,j}$ is then computed as:
\begin{equation}
	\begin{split}
		\begin{array}{rcl}
			\begin{aligned}
				\label{011_1}
				\ e_{k,j}=d_{k,j}-\hat{\bm m}^{\text T}_k\bm{x}_{k,j}=\big(\bm{m}_0-\hat{\bm m}_k\big)^{\text T}\bm{x}_{k,j}+v_{k,j}=\widetilde{\bm m}_k^{\text T}\bm{x}_{k,j}+v_{k,j},
			\end{aligned}
		\end{array}
	\end{split}
\end{equation}
where $\hat{\bm m}_k$ is the estimate of $\bm{m}_0$, $v_{k,j}$ represents subband noise signal, and $\widetilde{\bm{m}}_k\overset{\bigtriangleup}{=}\bm{m}_0-\hat{\bm m}_k$ denotes the weight deviation vector. The traditional NSAF algorithm updates the adaptive filter through \cite{1324714}
\begin{equation}
	\begin{split}
		\begin{array}{rcl}
			\begin{aligned}
				\label{011}
				 \hat{\bm m}_{k+r}=\hat{\bm m}_k+\mu\sum_{j=1}^N\frac{\bm{x}_{k,j}e_{k,j}}{\lvert\lvert \bm{x}_{k,j}\lvert\lvert_2^2}                            
			\end{aligned}
		\end{array}
	\end{split}
\end{equation}
using subband signals $\bm{x}_{k,j}$ and $e_{k,j}$, where $\mu$ is the step-size, $k=lr$, $l=0,1,...$, and the update interval $r$ adjusts the interval between iterations. However, when additive noise $v_{k,j}$ contains impulsive components, the outliers propagate into $e_{k,j}$ and destabilize the NSAF update process. To mitigate this, our prior work \cite{11113} introduced the NSPN algorithm, formulating a robust loss function inspired by the MPE criterion:
\begin{equation}
	\begin{split}
		\begin{array}{rcl}
			\begin{aligned}
				\label{011_2}
				J(k)=\sum_{j=1}^N \lvert e_{k,j}\lvert^p,
			\end{aligned}
		\end{array}
	\end{split}
\end{equation}
where $1\leq p<\alpha\leq2$. Using the stochastic gradient descent method, the NSPN algorithm can be formulated as
\begin{equation}
	\begin{split}
		\begin{array}{rcl}
			\begin{aligned}
				\label{011_3}
				\hat{\bm m}_{k+r}=\hat{\bm m}_k+\mu\sum_{j=1}^N\frac{\bm{x}_{k,j}\lvert e_{k,j}\lvert^{p-1} \text{sgn}(e_{k,j})}{\lvert\lvert \bm{x}_{k,j}\lvert\lvert_p^p},
			\end{aligned} 
		\end{array}
	\end{split}
\end{equation}
where $\text{sgn}(.)$ denotes the sign function and $\lvert\lvert\bm{x}_{k,j}\lvert\lvert_p^p=\sum_{i_1=0}^{D-1}\lvert x_{k-i_1,j}\lvert^p$ represents the $p$-norm of vector $\bm{x}_{k,j}$. 

\textbf{Remark 1}. When $p>1$, the cost function $J(k)$ in \eqref{011_2} is a strictly convex function with a unique minimum point \cite{339922}, which guarantees stable convergence of the NSPN algorithm under $\alpha$-stable noise ($\alpha \in (1,2]$). However, when $0 \leq p < \alpha \leq 1$, the cost function $J(k)$ is not first-order differentiable, and the stochastic gradient descent method cannot be used to minimize $J(k)$, causing significant performance degradation. Furthermore, non-Gaussian inputs introduce outliers that propagate through $\bm{x}_{k,j}$ in the weight update, generating erroneous gradients that destabilize the algorithm. Additionally,  for sparse systems where $\bm{m}_0$ exhibits structural sparsity, the NSPN framework cannot exploit this prior knowledge to accelerate convergence or reduce steady-state misadjustment.
\section{Proposed NKP-FoNSPN algorithms}
To address the limitations of the conventional NSPN algorithm outlined in Remark 1, this section introduces the NKP-FoNSPN and TNKP-FoNSPN algorithms. Furthermore, we establish the theoretical valid range for the fractional-order parameter $\beta$ and conduct a comprehensive computational complexity analysis comparing the proposed methods with state-of-the-art benchmarks.
\subsection{Derivation of NKP-FoNSPN algorithm}
Fig. 1 describes the structure of the NKP-FoNSPN algorithm, where the dashed lines denote the feedback paths for the subband error signals and the "Synthesizer" denotes the process of converting sub-filters $\hat{\bm m}_{k,2}$ and $\hat{\bm m}_{k,1}$ into the target filter $\hat{\bm m}_{k}$. We introduce the desired vector $\hat{\bm{d}}_{k} \in \mathbb{R}^{1\times L}$ and input matrix $\hat{\bm X}_k \in \mathbb{R}^{D\times L}$ as 
\begin{align}
	\label{013} 
	\hat{\bm{d}}_{k}\overset{\bigtriangleup}{=}[{d}_k, {d}_{k-1},...,{d}_{k-L+1}],
\end{align}
\vspace{-0.6cm}
\begin{align}
	\label{012} 
	\hat{\bm X}_{k}\overset{\bigtriangleup}{=}[\bm{x}_k,\bm{x}_{k-1},...,\bm{x}_{k-L+1}].
\end{align}

Then, we define the set formed by subband desired signals $\{{d}_{k,j}\}_{j=1}^N$ and subband input vectors $\{{\bm x}_{k,j}\}_{j=1}^N$ as the subband desired vector $\bm{d}_{k,s} \in \mathbb{R}^{1\times N}$ and subband input matrix $\bm{X}_{k,s} \in \mathbb{R}^{D\times N}$, respectively, i.e.,
\begin{align}
	\label{015} 
	\bm{d}_{k,s}\overset{\bigtriangleup}{=}[{d}_{k,1}, {d}_{k,2},..., {d}_{k,N}]=\hat{\bm{d}}_{k}\textbf{F},
\end{align}
\begin{align}
	\label{014a} 
	\bm{X}_{k,s}\overset{\bigtriangleup}{=}[\bm{x}_{k,1},\bm{x}_{k,2},...,\bm{x}_{k,N}]=\hat{\bm{X}}_{k}\textbf{F},
\end{align}
where $\textbf{F}\overset{\bigtriangleup}{=}[\bm{f}_1,\bm{f}_2,...,\bm{f}_N]$ denotes the analysis filter matrix with a size of $L\times N$. 
\begin{figure}
	\centering
	\includegraphics[scale=0.52] {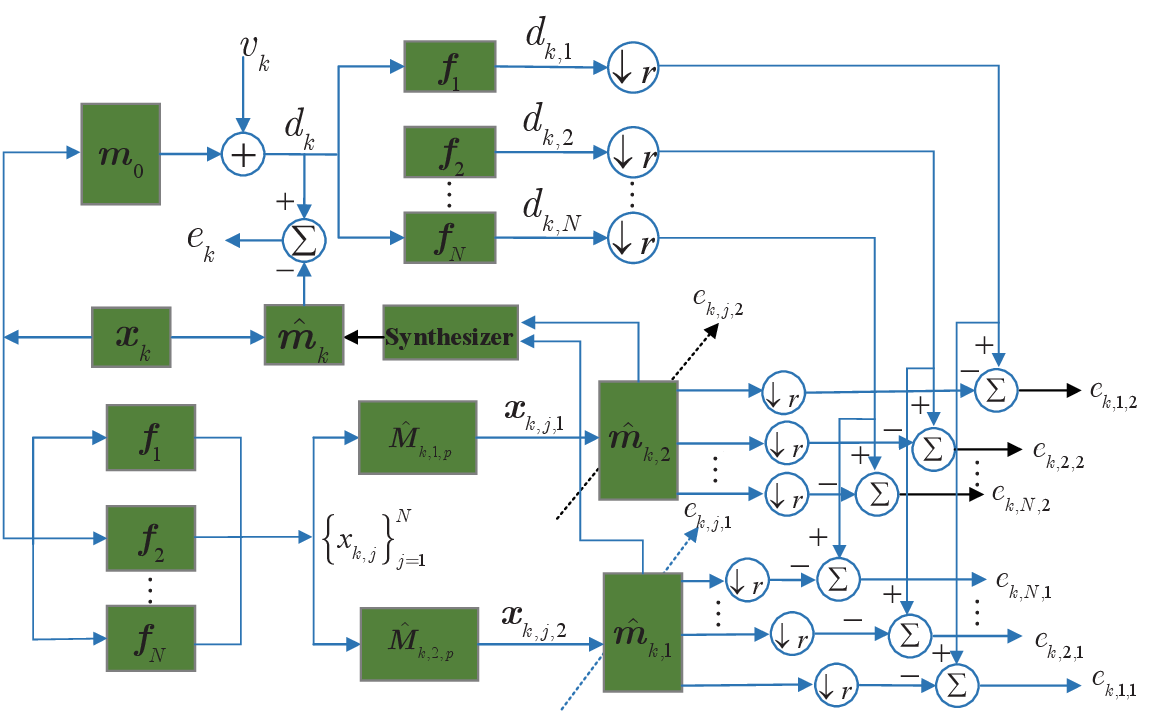}
	\vspace{-1em} \caption{Structure of NKP-FoNSPN.}
	\label{Fig0}
\end{figure}

Using a decomposition analogous to \eqref{010}, the adaptive filter $\hat{\bm m}_k$ is decomposed as
\begin{align}
	\label{016} 
	\hat{\bm m}_k=\sum_{q=1}^Q\hat{\bm m}_{k,2,q}\otimes\hat{\bm m}_{k,1,q},
\end{align}
where $\hat{\bm m}_{k,1,q} \in \mathbb{R}^{D_1\times 1}$ and $\hat{\bm m}_{k,2,q} \in \mathbb{R}^{D_2\times 1}$ represent the $q$-th column of sub-filter matrices $\widetilde{\bm M}_{k,1}\overset{\bigtriangleup}{=}[\hat{\bm m}_{k,1,1},\hat{\bm m}_{k,1,2},...,\hat{\bm m}_{k,1,Q}]\in \mathbb{R}^{D_1\times Q}$ and $\widetilde{\bm M}_{k,2}\overset{\bigtriangleup}{=}[\hat{\bm m}_{k,2,1},\hat{\bm m}_{k,2,2},...,\hat{\bm m}_{k,2,Q}]\in \mathbb{R}^{D_2\times Q }$, respectively. Furthermore, we define the following relationship
\begin{equation}
	\begin{split}
		\begin{array}{rcl}
			\begin{aligned}
				\label{017}
				\ \hat{\bm m}_{k,2,q}\otimes\hat{\bm m}_{k,1,q}&\overset{\bigtriangleup}{=}\big[\hat{\bm m}_{k,2,q}\otimes\bm{ I}_{D_1}\big]\hat{\bm m}_{k,1,q}\overset{\bigtriangleup}{=}\big[\bm{I}_{D_2}\otimes\hat{\bm m}_{k,1,q}\big]\hat{\bm m}_{k,2,q},
			\end{aligned}
		\end{array}
	\end{split}
\end{equation}
where $\bm{ I}_{D_1}$ and $\bm{ I}_{D_2}$ are identity matrices with sizes of $D_1\times D_1$ and $D_2\times D_2$, respectively.

By inserting \eqref{017} into \eqref{016}, we obtain
\begin{equation}
	\begin{split}
		\begin{array}{rcl}
			\begin{aligned}
				\label{018}
				\ \hat{\bm m}_k&=\sum_{q=1}^Q\hat{\bm M}_{k,2,q}\hat{\bm m}_{k,1,q}=\sum_{q=1}^Q\hat{\bm M}_{k,1,q}\hat{\bm m}_{k,2,q},
			\end{aligned}
		\end{array}
	\end{split}
\end{equation}
where $\hat{\bm M}_{k,1,q}\overset{\bigtriangleup}{=}\big[\bm{I}_{D_2}\otimes\hat{\bm m}_{k,1,q}\big]$ and $\hat{\bm M}_{k,2,q}\overset{\bigtriangleup}{=}\big[\hat{\bm m}_{k,2,q}\otimes\bm{ I}_{D_1}\big]$ of sizes $D_1D_2\times D_2$ and $D_1D_2\times D_1$, respectively.

By taking \eqref{018} into \eqref{011_1}, we obtain the subband error signals as
\begin{equation}
	\begin{split}
		\begin{array}{rcl}
			\begin{aligned}
				\label{020}
				\ e_{k,j,1}&=d_{k,j}-\sum_{q=1}^Q\hat{\bm m}_{k,1,q}^{\text T}\hat{\bm M}_{k,2,q}^{\text T}\bm{x}_{k,j}=d_{k,j}-\sum_{q=1}^Q\hat{\bm m}_{k,1,q}^{\text T}\bm{x}_{k,j,2,q}=d_{k,j}-\hat{\bm m}_{k,1}^{\text T}\bm{x}_{k,j,2},
			\end{aligned}
		\end{array}
	\end{split}
\end{equation}
\vspace{-0.6cm}
\begin{equation}
	\begin{split}
		\begin{array}{rcl}
			\begin{aligned}
				\label{021}
				\ e_{k,j,2}&=d_{k,j}-\sum_{q=1}^Q\hat{\bm m}_{k,2,q}^{\text T}\hat{\bm M}_{k,1,q}^{\text T}\bm{x}_{k,j}=d_{k,j}-\sum_{q=1}^Q\hat{\bm m}_{k,2,q}^{\text T}\bm{x}_{k,j,1,q}=d_{k,j}-\hat{\bm m}_{k,2}^{\text T}\bm{x}_{k,j,1},
			\end{aligned}
		\end{array}
	\end{split}
\end{equation}
where
\begin{equation}
	\begin{split}
		\begin{array}{rcl}
			\begin{aligned}
				\label{022}
				\ \bm{x}_{k,j,2,q}\overset{\bigtriangleup}{=}[{x}_{k,j,2,q},{x}_{k-1,j,2,q},...,{x}_{k-D_1+1,j,2,q}]^{\text T}=\hat{\bm M}_{k,2,q}^{\text T}\bm{x}_{k,j},
			\end{aligned}
		\end{array}
	\end{split}
\end{equation}
\vspace{-0.5cm} 
\begin{equation}
	\begin{split}
		\begin{array}{rcl}
			\begin{aligned}
				\label{023}
				\ \bm{x}_{k,j,2}=\big[\bm{x}_{k,j,2,1}^{\text T},\bm{x}_{k,j,2,2}^{\text T},...,\bm{x}_{k,j,2,Q}^{\text T}\big]^{\text T},
			\end{aligned}
		\end{array}
	\end{split}
\end{equation}
\vspace{-0.5cm} 
\begin{equation}
	\begin{split}
		\begin{array}{rcl}
			\begin{aligned}
				\label{023_1}
				\ \hat{\bm m}_{k,1}=[\hat{\bm m}_{k,1,1}^{\text T},\hat{\bm m}_{k,1,2}^{\text T},...,\hat{\bm m}_{k,1,Q}^{\text T} ]^{\text T},
			\end{aligned}
		\end{array}
	\end{split}
\end{equation}
\vspace{-0.7cm} 
\begin{equation}
	\begin{split}
		\begin{array}{rcl}
			\begin{aligned}
				\label{024}
				 \bm{x}_{k,j,1,q}\overset{\bigtriangleup}{=}[{x}_{k,j,1,q},{x}_{k-1,j,1,q},...,{x}_{k-D_2+1,j,1,q}]^{\text T}=\hat{\bm M}_{k,1,q}^{\text T}\bm{x}_{k,j},
			\end{aligned}
		\end{array}
	\end{split}
\end{equation}
\vspace{-0.5cm} 
\begin{equation}
	\begin{split}
		\begin{array}{rcl}
			\begin{aligned}
				\label{025}
				\ \bm{x}_{k,j,1}=\big[\bm{x}_{k,j,1,1}^{\text T},\bm{x}_{k,j,1,2}^{\text T},...,\bm{x}_{k,j,1,Q}^{\text T}\big]^{\text T},
			\end{aligned}
		\end{array}
	\end{split}
\end{equation}
\vspace{-0.5cm} 
\begin{equation}
	\begin{split}
		\begin{array}{rcl}
			\begin{aligned}
				\label{025_1}
				\ \hat{\bm m}_{k,2}=[\hat{\bm m}_{k,2,1}^{\text T},\hat{\bm m}_{k,2,2}^{\text T},...,\hat{\bm m}_{k,2,Q}^{\text T} ]^{\text T},
			\end{aligned}
		\end{array}
	\end{split}
\end{equation}
where the initial values of $\hat{\bm m}_{k,1,q}$ and $\hat{\bm m}_{k,2,q}$ are performed as $\hat{\bm m}_{0,1,q}=[\iota\;\bm{0}_{D_1-1}^{\text T}]^{\text T}$ and $\hat{\bm m}_{0,2,q}=[\iota\;\bm{0}_{D_2-1}^{\text T}]^{\text T}$ with $0<\iota\leq 1$ \cite{9053421,8682498}, we refer to this method as Method-I. Of course, we can also set the initial values of $\hat{\bm m}_{k,1,q}$ and $\hat{\bm m}_{k,2,q}$ to be $\hat{\bm m}_{0,1,q}=\iota\bm{I}_{D_1\times 1}$ and $\hat{\bm m}_{0,2,q}=\iota\bm{I}_{D_2\times 1}$, respectively, we refer to this method as Method-II. To the best of our knowledge, at present, there is no literature that specifies the method for selecting parameter $\iota$.

To construct robust weight update rules for the sub-filters $\hat{\bm m}_{k,1}$ and $\hat{\bm m}_{k,2}$, we formulate the MPE-inspired loss functions as
\begin{equation}
	\begin{split}
		\begin{array}{rcl}
			\begin{aligned}
				\label{026}
				J_1(k)=\sum_{j=1}^N \lvert e_{k,j,1}\lvert^p,
			\end{aligned}
		\end{array}
	\end{split}
\end{equation}
\vspace{-0.5cm}
\begin{equation}
	\begin{split}
		\begin{array}{rcl}
			\begin{aligned}
				\label{026_1}
				J_2(k)=\sum_{j=1}^N \lvert e_{k,j,2}\lvert^p,
			\end{aligned}
		\end{array}
	\end{split}
\end{equation}
and then we employ the FoSGD criterion to derive the $\beta$-order gradient-based update rules for the sub-filters $\hat{\bm m}_{k,1}$ and $\hat{\bm m}_{k,2}$ as:
\begin{equation}
	\begin{split}
		\begin{array}{rcl}
			\begin{aligned}
				\label{026_2}
				\hat{\bm m}_{k+r,1} = 	\hat{\bm m}_{k,1} -\mu \nabla^{\beta}J_1(k),
			\end{aligned}
		\end{array}
	\end{split}
\end{equation}
\vspace{-0.5cm}
\begin{equation}
	\begin{split}
		\begin{array}{rcl}
			\begin{aligned}
				\label{026_3}
				\hat{\bm m}_{k+r,2} = 	\hat{\bm m}_{k,2} -\mu \nabla^{\beta}J_2(k),
			\end{aligned}
		\end{array}
	\end{split}
\end{equation}
where $\nabla^{\beta}(.)$ denotes $\beta$-order gradient operator.

Given that the sub-filter structures after decomposition are similar, we derive only the weight update formula for filter $\hat{\bm m}_{k,1}$ to avoid redundancy in the derivation process. To facilitate the subsequent derivation, we introduce the following lemmas. 

\textbf{Lemma 1:} \textit{The fractional derivative of any compound function } $G\big(y(x)\big)$ \textit{follows the chain rule defined as \cite{jumarie2013}}
\begin{equation}
	\begin{split}
		\begin{array}{rcl}
			\begin{aligned}
				\label{027}
				\Big[{G}\big(y(x)\big)\Big]^{(\beta)}=G_y^{(\beta)}(y)(y_x^{'})^{\beta},
			\end{aligned}
		\end{array}
	\end{split}
\end{equation}
where $\beta$ represents the order of the fractional derivative, and $y_x^{'}$ denotes the first-order derivative with respect to $x$.

\textbf{Lemma 2:} \textit{ For the fractional derivative of any power function $y(x)=(x-c)^p$, the following equality holds \cite{jumarie2013}:} 
\begin{equation}
	\begin{split}
		\begin{array}{rcl}
			\begin{aligned}
				\label{028}
			y_x^{(\beta)}(x)=\Gamma(p+1)\Gamma^{-1}(p+1-\beta)(x-c)^{p-\beta},\;\; n>0,
			\end{aligned}
		\end{array}
	\end{split}
\end{equation}
where $\beta<p$.

By utilizing the fractional-order chain rule defined in \textbf{Lemma 1}, the term $\nabla^{\beta}J_1(k)$ in \eqref{026_2} can be calculated by
\begin{equation}
	\begin{split}
		\begin{array}{rcl}
			\begin{aligned}
				\label{029}
				\nabla^{\beta}J_1(k)=J_{e_{k,j,1}}^{(\beta)}(k)\Bigg[\frac{\partial\big( e_{k,j,1}\big)}{\partial \hat{\bm m}_{k,1}}\Bigg]^{\beta}.
			\end{aligned}
		\end{array}
	\end{split}
\end{equation}
Then, based on \textbf{Lemma 2}, the first part in the right-hand side of \eqref{029} can be derived as
\begin{equation}
	\begin{split}
		\begin{array}{rcl}
			\begin{aligned}
				\label{030}
				J_{e_{k,j,1}}^{(\beta)}(k)=\sum_{j=1}^N\frac{\Gamma(p+1)}{\Gamma(p-\beta+1)}\lvert e_{k,j,1}\lvert^{p-\beta}\text{sgn}(e_{k,j,1}),
			\end{aligned}
		\end{array}
	\end{split}
\end{equation}
and the second part in the right-hand side of \eqref{029} can be derived as 
\begin{equation}
	\begin{split}
		\begin{array}{rcl}
			\begin{aligned}
				\label{031}
				\Bigg[\frac{\partial\big( e_{k,j,1}\big)}{\partial \hat{\bm m}_{k,1}}\Bigg]^{\beta}=(-\bm{x}_{k,j,2})^{\beta},
			\end{aligned}
		\end{array}
	\end{split}
\end{equation}
where $(-\bm{x}_{k,j,2})^{\beta}=\big[(-{x}_{k,j,2,1})^{\beta},(-{x}_{k-1,j,2,1})^{\beta},...,(-{x}_{k-D_1+1,j,2,1})^{\beta},...,(-{x}_{k,j,2,Q})^{\beta},(-{x}_{k-1,j,2,Q})^{\beta},...,(-{x}_{k-D_1+1,j,2,Q})^{\beta}\big]^{\text T}$.

The last term in \eqref{031} may generate complex values if $\bm{x}_{k,j,2}$ has positive elements. In many real-valued signal processing scenarios, the imaginary part of the result can be ignored. Therefore, \eqref{031} can be approximated as
\begin{equation}
	\begin{split}
		\begin{array}{rcl}
			\begin{aligned}
				\label{035}
				\Bigg[\frac{\partial\big( e_{k,j,1}\big)}{\partial \hat{\bm m}_{k,1}}\Bigg]^{\beta}\approx \text{Re}\Big\{(-\bm{x}_{k,j,2})^{\beta}\Big\}.
			\end{aligned}
		\end{array}
	\end{split}
\end{equation}

Substituting \eqref{029}, \eqref{030}, and \eqref{035} into \eqref{026_2}, we obtain
\begin{equation}
	\begin{split}
		\begin{array}{rcl}
			\begin{aligned}
				\label{036}
				\hat{\bm m}_{k+r,1}=\hat{\bm m}_{k,1}-\mu \sum_{j=1}^N g(e_{k,j,1})\text{Re}\big\{(-\bm{x}_{k,j,2})^{\beta}\big\},
			\end{aligned}
		\end{array}
	\end{split}
\end{equation}
where $g(e_{k,j,1})=\text{sgn}(e_{k,j,1})\lvert e_{k,j,1}\lvert^{p-\beta}$ denotes the nonlinear error term, and the term $\Gamma(p+1)/\Gamma(p-\beta+1)$ is absorbed into the step-size $\mu$.

To utilize the subband adaptive filtering's decorrelation for correlated input signals, we introduce the normalization term in \eqref{036} and obtain the weight update formula of sub-filter $\hat{\bm m}_{k,1}$, i.e., 
\begin{equation}
	\begin{split}
		\begin{array}{rcl}
			\begin{aligned}
				\label{037}
				\hat{\bm m}_{k+r,1}=\hat{\bm m}_{k,1}-\mu \sum_{j=1}^N \frac{g(e_{k,j,1})\text{Re}\big\{(-\bm{x}_{k,j,2})^{\beta}\big\}}{\lvert\lvert \bm{x}_{k,j,2}\lvert\lvert_p^p},
			\end{aligned}
		\end{array}
	\end{split}
\end{equation}
where $\lvert\lvert \bm{x}_{k,j,2}\lvert\lvert_p^p=\sum_{q=1}^{Q}\sum_{i_2=0}^{D_1-1}\lvert x_{k-i_2,j,2,q}\lvert^p$.

Similarly, the weight update formula of sub-filter $\hat{\bm m}_{k,2}$ can be described as
\begin{equation}
	\begin{split}
		\begin{array}{rcl}
			\begin{aligned}
				\label{038}
				\hat{\bm m}_{k+r,2}=\hat{\bm m}_{k,2}-\mu \sum_{j=1}^N \frac{g(e_{k,j,2})\text{Re}\big\{(-\bm{x}_{k,j,1})^{\beta}\big\}}{\lvert\lvert \bm{x}_{k,j,1}\lvert\lvert_p^p},
			\end{aligned}
		\end{array}
	\end{split}
\end{equation}
where $g(e_{k,j,2})=\text{sgn}(e_{k,j,2})\lvert e_{k,j,2}\lvert^{p-\beta}$ represents the nonlinear error term. 

As a summary, \eqref{016}, \eqref{020}, \eqref{021}, \eqref{023}, \eqref{025}, \eqref{037}, and \eqref{038} constitute the proposed NKP-FoNSPN algorithm. Notably, the normalization terms $1/\lvert\lvert \bm{x}_{k,j,1}\lvert\lvert_p^p$ and $1/\lvert\lvert \bm{x}_{k,j,2}\lvert\lvert_p^p$ can also be utilized to adjust the step-size. When the subband input signals $\{\bm{x}_{k,j,1}\}_{j=1}^N$ and $\{\bm{x}_{k,j,2}\}_{j=1}^N$ experience impulsive interference, this mechanism selects a smaller step-size to enhance robustness against impulsive noise.

\textbf{Remark 2}. When $\beta=1$, the proposed NKP-FoNSPN algorithm degenerates into a novel NKP-NSPN algorithm, which has good performance in the $\alpha$-stable noise ($1\leq p<\alpha\leq2$) and sparse SI. The NKP-NSPN algorithm is formulated as
\begin{equation}
	\begin{split}
		\begin{array}{rcl}
			\begin{aligned}
				\label{043}
				\hat{\bm m}_{k+r,1}=\hat{\bm m}_{k,1}-\mu \sum_{j=1}^N \frac{\lvert e_{k,j,1}\lvert^{p-2}e_{k,j,1}\text{Re}\big\{(-\bm{x}_{k,j,2})\big\}}{\lvert\lvert \bm{x}_{k,j,2}\lvert\lvert_p^p},
			\end{aligned}
		\end{array}
	\end{split}
\end{equation}
\vspace{-0.3cm}
\begin{equation}
	\begin{split}
		\begin{array}{rcl}
			\begin{aligned}
				\label{044}
				\hat{\bm m}_{k+r,2}=\hat{\bm m}_{k,2}-\mu \sum_{j=1}^N \frac{\lvert e_{k,j,2}\lvert^{p-2}e_{k,j,2}\text{Re}\big\{(-\bm{x}_{k,j,1})\big\}}{\lvert\lvert \bm{x}_{k,j,1}\lvert\lvert_p^p}.
			\end{aligned}
		\end{array}
	\end{split}
\end{equation}

\textbf{Remark 3}. Non-NKP decomposition version of the proposed NKP-FoNSPN algorithm is the FoNSPN algorithm, i.e.,
\begin{equation}
	\begin{split}
		\begin{array}{rcl}
			\begin{aligned}
				\label{045}
				\hat{\bm m}_{k+r}=\hat{\bm m}_{k}-\mu \sum_{j=1}^N \frac{g(e_{k,j})\text{Re}\big\{(-\bm{x}_{k,j})^{\beta}\big\}}{\lvert\lvert \bm{x}_{k,j}\lvert\lvert_p^p},
			\end{aligned}
		\end{array}
	\end{split}
\end{equation}
which has good performance in both non-Gaussian inputs and $\alpha$-stable noise with $0\leq p<\alpha\leq1$. Note that, when $\beta=1$, the FoNSPN algorithm degenerates into the conventional NSPN algorithm.
\subsection{Derivation of TNKP-FoNSPN algorithm}
It is noteworthy that through the exploration of existing NKP decomposition-based adaptive filtering algorithms, we have observed that the NKP decomposition technology significantly enhances the convergence performance. However, step-size adjustment alone cannot enable the NKP decomposition-based algorithms to achieve the steady-state performance comparable to that of the non-NKP decomposition algorithms with a small step-size. To facilitate understanding, we designed the experimental scheme shown in Fig. 2 using the NKP-NLMS algorithm as an example.
\begin{figure}
	\centering
	\includegraphics[scale=0.45] {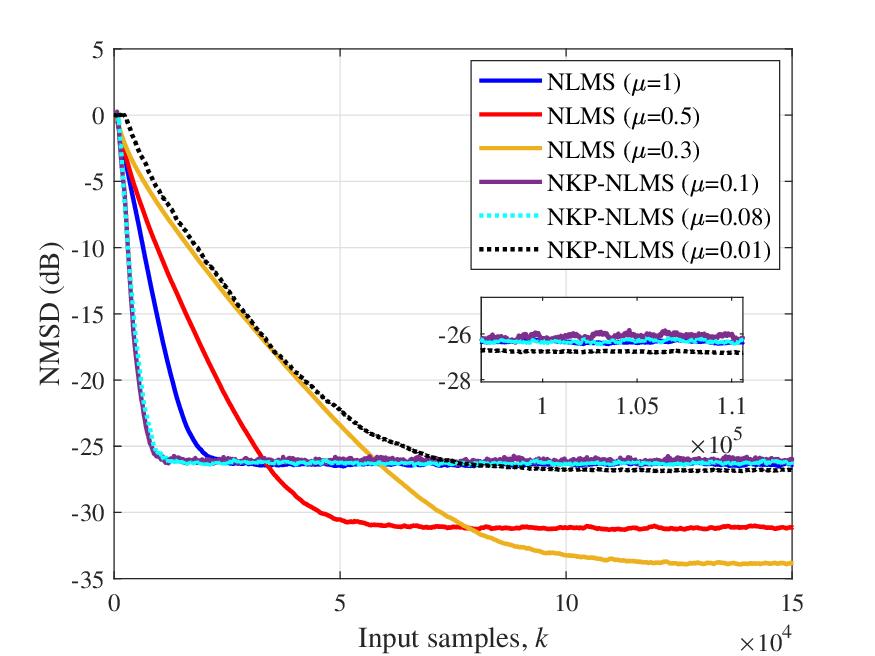}
	\vspace{-1em} \caption{NMSD curves of the NKP-NLMS and NLMS algorithms.}
	\label{Fig1}
\end{figure}

As shown in Fig. 2, the NKP-NLMS algorithm achieves faster convergence than the NLMS algorithm with $\mu$=1 (which is its optimal step-size for maximizing convergence rate) during the transient stage. However, step-size adjustment alone cannot enable the NKP-NLMS algorithm to reach the same steady-state misadjustment level as the NLMS algorithm with $\mu$=0.3. This limitation also persists in the proposed NKP-FoNSPN algorithm. To simultaneously retain the rapid transient convergence of NKP-based algorithms and the low steady-state misadjustment of non-NKP-based approaches, we propose a novel TNKP decomposition strategy presented in Table 2. Clealy, during the transient stage, the TNKP-based algorithm updates according to the NKP-based algorithm to achieve faster convergence rate. Conversely, during the steady-state stage, the TNKP-based algorithm updates according to the non-NKP algorithm to achieve lower steady-state misadjustment. The boundary between the transient and steady-state stages is controlled by the parameter $\rho$, which is calculated by taking the mean of last five thousand steady-state NMSD of the NKP-based algorithm. Notably, the TNKP strategy applies not only to NKP-FoNSPN but also to other NKP-based adaptive filtering algorithms. For specific results, please refer to the simulation section.
\begin{table}
	\renewcommand\arraystretch{1}
	\scriptsize
	\centering
	\caption{Proposed transformation NKP decomposition strategy}
	\label{table_2}
	\begin{tabular}{lc}
		\hline
		\text{Initialization:} $\text{flag}=0$, $\text{Flag}=0$;\\ 
		\hline
		\text{if} $\text{flag}=0$\\
		\;\;\;\;\;$\hat{\bm m}_{k+r,1}=\hat{\bm m}_{k,1}-\mu \sum_{j=1}^N \frac{g(e_{k,j,1})\text{Re}\big\{(-\bm{x}_{k,j,2})^{\beta}\big\}}{\lvert\lvert \bm{x}_{k,j,2}\lvert\lvert_p^p}$;\\
		\;\;\;\;\;$\hat{\bm m}_{k+r,2}=\hat{\bm m}_{k,2}-\mu \sum_{j=1}^N \frac{g(e_{k,j,2})\text{Re}\big\{(-\bm{x}_{k,j,1})^{\beta}\big\}}{\lvert\lvert \bm{x}_{k,j,1}\lvert\lvert_p^p}$;\\
		\;\;\;\;\;$\hat{\bm m}_k=\sum_{q=1}^Q\hat{\bm m}_{k,2,q}\otimes\hat{\bm m}_{k,1,q}$;\\
		\text{else}\\
		\;\;\;\;\;$\hat{\bm m}_{k+r}=\hat{\bm m}_{k}-\mu_b \sum_{j=1}^N \frac{g(e_{k,j})\text{Re}\big\{(-\bm{x}_{k,j})^{\beta}\big\}}{\lvert\lvert \bm{x}_{k,j}\lvert\lvert_p^p}$;\\
		end\\
		\text{if} $\text{Flag}=0$\\
		\;\;\;\text{if} $\text{NMSD}_k \leq \rho$\\
		\;\;\;\;\;\;\;$\text{flag}=1$;\\
		\;\;\;\;\;\;\;$\text{Flag}=1$;\\
		\;\;\;\text{else}\\
		\;\;\;\;\;\;\;$\text{flag}=0$;\\
		\;\;\;end\\
		end\\
		\hline
	\end{tabular}
\end{table} 
\subsection{Bound on the Fractional-Order $\beta$}
When the unknown system $\bm{m}_0$ is a sparse system with a low-rank structure, choosing $Q$ to be close to the true rank can minimize the NKP decomposition error \cite{10979895}. Given that the sub-filter structures after decomposition are similar, we analyze the convergence range of the global parameter $\beta$ by taking one of the sub-filters as an example. To proceed, by defining the increment $H_k\overset{\bigtriangleup}{=}\hat{\bm m}_{k+r,1}-\hat{\bm m}_{k,1}$, and the expectation of Euclidean norm of \eqref{037} can be formulated as 
\begin{equation}
	\begin{split}
		\begin{array}{rcl}
			\begin{aligned}
				\label{046} 
				&\text{E}\big\{\lvert\lvert H_k\lvert\lvert_2^2\big\}=\mu^2\sum_{j=1}^N \text{E}\Big\{\frac{\lvert e_{k,j,1}\lvert^{2(p-\beta)}\lvert\lvert \text{Re}\{(-\bm{x}_{k,j,2})^{\beta}\}\lvert\lvert_2^2}{\lvert\lvert \bm{x}_{k,j,2}\lvert\lvert_p^{2p}}\Big\} +\mu^2 \sum_{j=1\atop {j\neq t}}^N \text{E}\Big\{\frac{g(e_{k,j,2})\text{Re}\big\{(-\bm{x}_{k,j,2})^{\beta}\big\}}{\lvert\lvert \bm{x}_{k,j,2}\lvert\lvert_p^p} \frac{g(e_{k,t,2})\text{Re}\big\{(-\bm{x}_{k,t,2})^{\beta}\big\}}{\lvert\lvert \bm{x}_{k,t,2}\lvert\lvert_p^p}\Big\} .
			\end{aligned}
		\end{array}
	\end{split}
\end{equation}
Since the assumption that the weak correlation between the input signals $\{\bm{x}_{k,j,1}\}_{j=1}^N$ at different subbands \cite{1324714}, the final part on the right-hand side of \eqref{046} can be
neglected. In addition, since $e_{k,j,1}$ is statistically independent of $\bm{x}_{k,j,2}$ in the steady-state stage \cite{al2001adaptive}, \eqref{046} can be simplified to
\begin{equation}
	\begin{split}
		\begin{array}{rcl}
			\begin{aligned}
				\label{047} 
				&\text{E}\big\{\lvert\lvert H_k\lvert\lvert_2^2\big\}=\mu^2\sum_{j=1}^N \text{E}\big\{\lvert e_{k,j,1}\lvert^{2(p-\beta)}\big\} \text{E}\Big\{\frac{\lvert\lvert \text{Re}\{(-\bm{x}_{k,j,2})^{\beta}\}\lvert\lvert_2^2}{\lvert\lvert \bm{x}_{k,j,2}\lvert\lvert_p^{2p}}\Big\} .
			\end{aligned}
		\end{array}
	\end{split}
\end{equation}

Since the influence exerted by $\text{E}\Big\{\frac{\lvert\lvert \text{Re}\{(-\bm{x}_{k,j,2})^{\beta}\}\lvert\lvert_2^2}{\lvert\lvert \bm{x}_{k,j,2}\lvert\lvert_p^{2p}}\Big\}$ on $\text{E}\big\{\lvert\lvert H_k\lvert\lvert_2^2\big\}$ is limited \cite{3}. In other words, the value of $\text{E}\big\{\lvert\lvert H_k\lvert\lvert_2^2\big\}$ mainly depends on $\mu^2\sum_{j=1}^N \text{E}\big\{\lvert e_{k,j,1}\lvert^{2(p-\beta)}\big\}$. Thus, when NKP-FoNSPN algorithm converges to steady-state stage, $\text{E}\big\{\lvert\lvert H_k\lvert\lvert_2^2\big\} \rightarrow 0$ is equivalent to $\mu^2\sum_{j=1}^N \text{E}\big\{\lvert e_{k,j,1}\lvert^{2(p-\beta)}\big\}  \rightarrow 0$.

Based on the existence condition $0\leq p<\alpha$ for fractional-order lower-order moments, to ensure the finiteness of the expectation $\text{E}\big\{\lvert e_{k,j,1}\lvert^{2(p-\beta)}\big\}$, the exponent $2(p-\beta)$ must satisfy this existence condition. Thus, we obtain
\vspace{-0.1cm} 
\begin{equation}
	\begin{split}
		\begin{array}{rcl}
			\begin{aligned}
				\label{049} 
				0\leq 2(p-\beta)<\alpha,
			\end{aligned}
		\end{array}
	\end{split}
\end{equation}
which is equivalent to
\begin{equation}
	\begin{split}
		\begin{array}{rcl}
			\begin{aligned}
				\label{050} 
				p-\frac{\alpha}{2}<\beta\leq p.
			\end{aligned}
		\end{array}
	\end{split}
\end{equation}
This conclusion is further verified in Fig. 7 of Section 5.
\subsection{Analysis of computational complexity}
Table 3 summarizes the computational complexity of the NSPN, FoNSPN, FoNLMP, FoMVC, NKP-GHSAF, NKP-RLS, NKP-NSPN, and NKP-FoNSPN algorithms in terms of multiplication, addition, power, and linear operations for every $r$ input samples. "Power" denotes the calculation of exponential and power functions, while "Sign" refers to the symbolic function, absolute value, and real part operations. It is worth noting that subband-based adaptive filtering algorithms require $LN(D+1)$ multiplications and $(LN-N)(D+1)$ additions to decompose the input vector $\bm{x}_k$ and desired signal $d_k$ into subband input vectors $\{\bm{x}_{k,j}\}_{j=1}^N$ and subband desired signals $\{d_{k,j}\}_{j=1}^N$ using analysis filters, respectively. To enable a more intuitive comparison of computational costs, Fig. 3 contrasts the influence of varying adaptive filter lengths $D$ and parameters $Q$ on the multiplication and power operation requirements of the algorithms.

\textbf{Remark 4.} According to the implementation principle of the TNKP decomposition technology, the computational complexity of the TNKP-FoNSPN algorithm lies between that of the NKP-FoNSPN and FoNSPN algorithms throughout the adaptive stage. Furthermore, since the NKP-FoNSPN algorithm inherits the fast convergence capability of the NKP decomposition technology, the TNKP-FoNSPN algorithm exhibits computational complexity closer to that of the FoNSPN algorithm.

Figs. 3 (a) and (b) illustrate the computational complexity of the NSPN, FoNMLP, FoNSPN, FoMVC, NKP-GHSAF, NKP-RLS, NKP-NSPN, and NKP-FoNSPN algorithms with respect to parameter $Q$. The NKP-RLS algorithm involves matrix operations, resulting in significantly higher multiplication costs than other algorithms. The multiplication cost per $r$ input samples is identical for the NKP-FoNSPN and NKP-NSPN algorithms. However, the former sacrifices minimal power operation cost to achieve superior performance in handling non-Gaussian inputs and $\alpha$-stable noise ($0<\alpha \leq1$). Although the proposed NKP-FoNSPN algorithm exhibits higher multiplication and power costs than NKP-GHSAF, the latter demonstrates inferior convergence rates and reduced robustness to non-Gaussian inputs and $\alpha$-stable noise ($0<\alpha \leq1$). Moreover, the multiplication cost gap between NKP-FoNSPN and the FoNMLP, FoMVC, NSPN, and FoNSPN algorithms widens with increasing $Q$. However, for $Q<5$, NKP-FoNSPN maintains lower power operation costs than these algorithms, while its enhanced convergence and steady-state performance provide compensatory benefits relative to computational complexity.
\begin{figure}
	\centering  
	\includegraphics[scale=0.48] {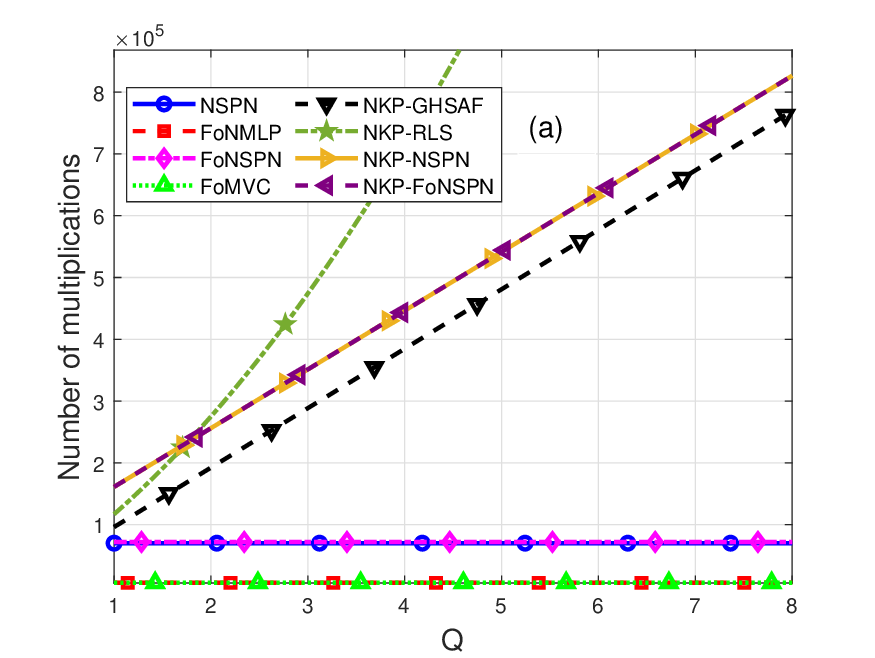}  
	\hspace{0.1ex}	
	\includegraphics[scale=0.48] {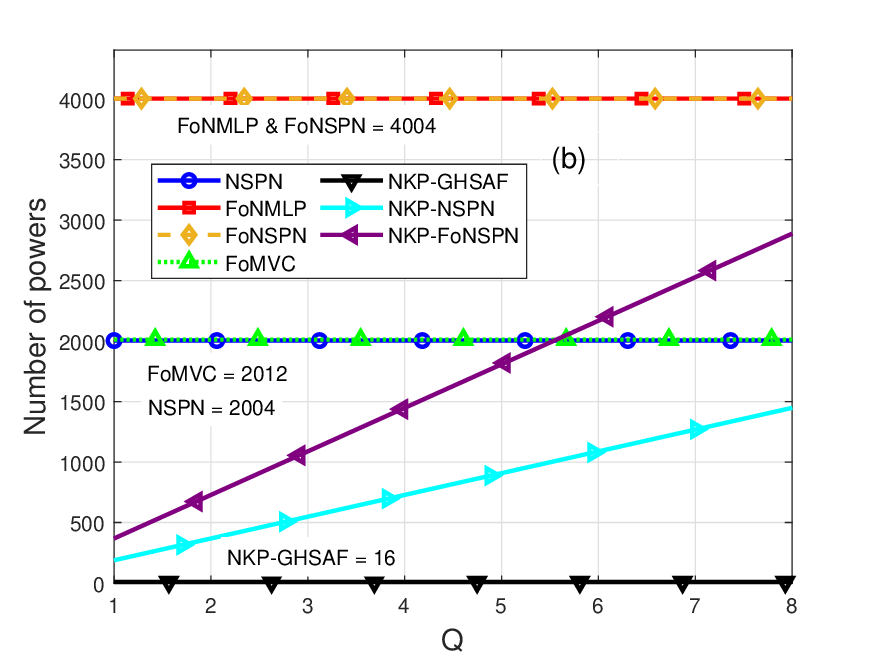}
	\hspace{0.1ex}						 
	\includegraphics[scale=0.48] {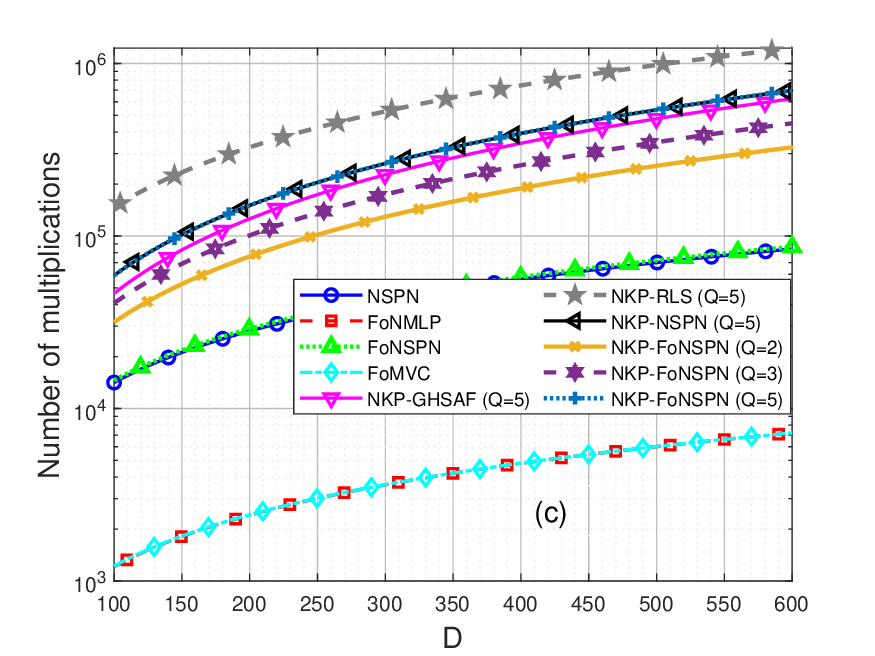}	
	\hspace{0.1ex}									 
	\includegraphics[scale=0.48] {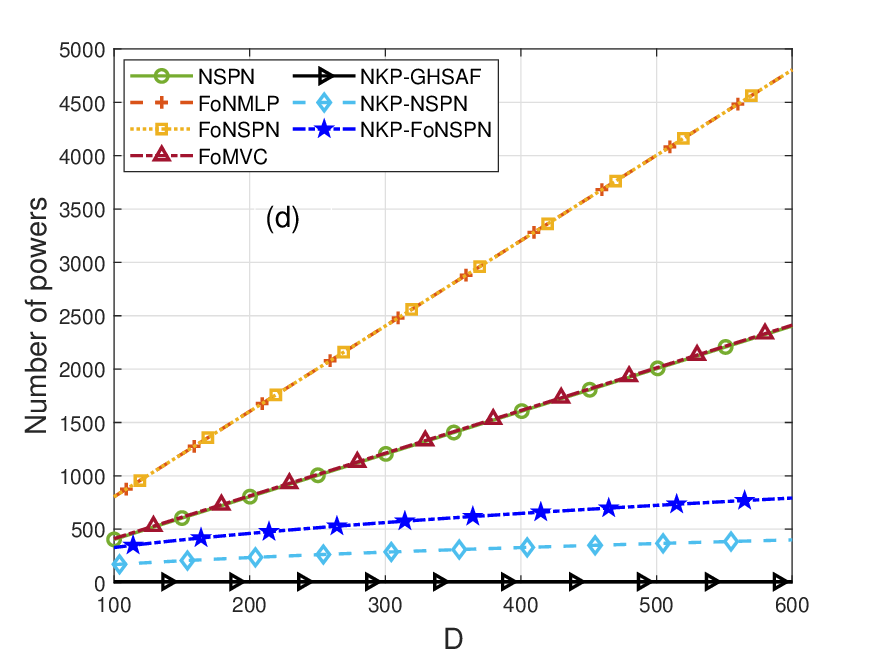}						  
	\caption{Computational complexity of the NSPN \cite{11113}, FoNLMP \cite{3}, FoNSPN, FoMVC \cite{8}, NKP-GHSAF \cite{9170797}, NKP-RLS \cite{8682498}, NKP-NSPN, and NKP-FoNSPN algorithms. (a) The computational complexity of multiplications versus the parameter $Q$; (b) the computational complexity of powers versus the parameter $Q$; (c) the computational complexity of multiplications versus the parameter $D$; (d) the computational complexity of powers versus the parameter $D$. $D=500$, $D_1=25$, and $D_2=20$ for Figs. 3 (a) and (b); $D_1$=$D_2$=$\sqrt{D}$ for Figs. 3 (c) and (d); $Q=2$ for Fig. 3 (d). $L=33$, $N=4$, $r=4$. }
	\label{Figq}
\end{figure}

Fig. 3 (c) illustrates the computational complexity of the NSPN, FoNMLP, FoNSPN, FoMVC, NKP-GHSAF, NKP-RLS, NKP-NSPN, and NKP-FoNSPN algorithms versus the parameter $D$. The multiplication cost of all algorithms increases with $D$, while NKP decomposition-based adaptive filtering algorithms exhibit higher multiplication costs than their non-NKP-based counterparts. Furthermore, the computational complexity of the NKP-FoNSPN algorithm is marginally higher than that of NKP-GHSAF but lower than  that of NKP-RLS. Fig. 3(d) shows that the power operation cost of the proposed NKP-NSPN and NKP-FoNSPN algorithms is slightly elevated compared to NKP-GHSAF but substantially lower than that of the four non-NKP-based algorithms (NSPN, FoNMLP, FoNSPN, and FoMVC).

\section{Proposed filtered-x NKP-FoNSPN algorithms }
\begin{table*}[htp]
	\renewcommand\arraystretch{1.2}
	\tabcolsep = 0.07cm
	
	\setlength{\abovecaptionskip}{0cm}
	\setlength{\belowcaptionskip}{0cm}
	\begin{center}
		\caption{ Computational complexity of different algorithms for every $r$ input samples  in system identification}
		\begin{tabular}{c c c c c } 
			\hline
			\rowcolor{gray!30}	\footnotesize{Algorithms} &\footnotesize{$\times$/$\div$}  &\footnotesize{$+$/$-$}  &\footnotesize{Power} & \footnotesize{Linear}   \\			
			\hline
			\footnotesize{NSPN}	 &\footnotesize{$LN(D$+$1)$+$2DN$+$3N$}	&\footnotesize{$(LN$-$N)(D$+$1)$+$4D$+$2DN$}&\footnotesize{$(D$+$1)N$} & \footnotesize{$(D$+$2)N$ }\\
			\rowcolor{gray!30}\footnotesize{FoNSPN} &\footnotesize{$LN(D$+$1)$+$3DN$+$3N$}&\footnotesize{$(LN$-$N)(D$+$1)$+$4D$+$2DN$}&\footnotesize{$(2D$+$1)N$ }& \footnotesize{$(2D$+$2)N$}\\
			\footnotesize{FoNMLP} &\footnotesize{$(3D$+$3)r$}& \footnotesize{$(3D$+$1)r$}& \footnotesize{$(2D$+$1)r$}&  \footnotesize{$(2D$+$1)r$}\\
			\rowcolor{gray!30}\footnotesize{FoMVC}&\footnotesize{$(3D$+$5)r$ }& \footnotesize{$(2D$+$3)r$} & \footnotesize{$(D$+$3)r$}& \footnotesize{$(2D$+$3)r$}\\
			\footnotesize{NKP-GHSAF} &\footnotesize{$\big(3DQ$+$(D_1$+$D_2)DQ$}&\footnotesize{$\big((DQ$-$Q)(D_1$+$D_2)$+$2$}&\footnotesize{$4r$ }& \footnotesize{$6r$}\\
			& \footnotesize{$+(D_1$+$D_2)Q$+$10\big)r$}&\footnotesize{$+(Q$-$1)D$+$(D_1$+$D_2)Q\big)r$}&& \\
			\rowcolor{gray!30}\footnotesize{NKP-RLS} &\footnotesize{$\big(5Q^2(D_1^2$+$D_2^2)$+$3Q(D_1$+$D_2)$}&\footnotesize{$3Q^2(D_1^2$+$D_2^2)$+$(Q-1)D$} &\footnotesize{0}& \footnotesize{0}\\
			\rowcolor{gray!30}&\footnotesize{$+QD(D_1$+$D_2$+$3)\big)r$}&\footnotesize{$+(D_1+D_2)(DQ-Q)-2\big)r$}& & \\
			\footnotesize{NKP-NSPN} &\footnotesize{$DQ$+$6N$+$3(D_1$+$D_2)NQ$+$2DQN$}&\footnotesize{$(LN$-$N)(1$+$D)$+$(DNQ$-$NQ)(D_1$+$D_2)$}&\footnotesize{$2N$+$(D_1$+$D_2)NQ$ }&\footnotesize{$2N$+$2(D_1$+$D_2)NQ$}\\
			&\footnotesize{$+(D$+$1)LN$+$(D_1$+$D_2)DQN$}&\footnotesize{$+(Q$-$1)D$+$(2QN$+$4Q)(D_1$+$D_2)$}& &\\
			\rowcolor{gray!30}	\footnotesize{NKP-FoNSPN} &\footnotesize{$DQ$+$6N$+$3(D_1$+$D_2)NQ$+$2DQN$}&\footnotesize{$(LN$-$N)(1$+$D)$+$(DNQ$-$NQ)(D_1$+$D_2)$}&\footnotesize{$2N$+$2(D_1$+$D_2)NQ$ }&\footnotesize{$2N$+$2(D_1$+$D_2)NQ$}\\
			\rowcolor{gray!30}	&\footnotesize{$+(D$+$1)LN$+$(D_1$+$D_2)DQN$}&\footnotesize{$+(Q$-$1)D$+$(2QN$+$4Q)(D_1$+$D_2)$}& & \\
			\hline
		\end{tabular}
	\end{center}
\end{table*}
In this section, we construct the ANC models for the aforementioned proposed algorithms to address increasingly complex noise environments. Fig. 4 illustrates the structure of the NKP-FxFoNSPN algorithm, where the dashed lines denote unknown paths that need adaptive estimation. The equivalent discrete transfer function $P(z)$ of the primary path is used to model the acoustic echo path between the reference microphone and the error microphone, while $S(z)$ models the electro-acoustic path from the secondary loudspeaker to the error microphone. In practical ANC scenarios, $S(z)$ is unknown but can be estimated as $\hat{S}(z)$ through offline or online modeling techniques. The signal $\bm{x}_k$ captured by the reference microphone denotes the noise source, also known as the reference signal, which propagates through $P(z)$ to generate the desired signal $d_k$ (primary disturbance). Concurrently, $\bm{x}_k$ is processed by the adaptive filter $\hat{\bm m}_k$, with the output convolved with the secondary path $S(z)$ to produce the anti-noise signal $y_k$. The residual error $e_k$ is then obtained by subtracting $y_k$ from $d_k$:
\begin{equation}
	\begin{split}
		\begin{array}{rcl}
			\begin{aligned}
				\label{051} 
			e_k=d_k-y_k=d_k-\hat{s}(k)* \big({\bm x}_k^{\text T}\hat{\bm m}_k\big),
			\end{aligned}
		\end{array}
	\end{split}
\end{equation}
where "$*$" represents the linear convolution operator, $\hat{s}(k)$ is the IR of the secondary path $\hat{S}(z)$. 
\begin{figure}
	\centering
	\includegraphics[scale=0.48] {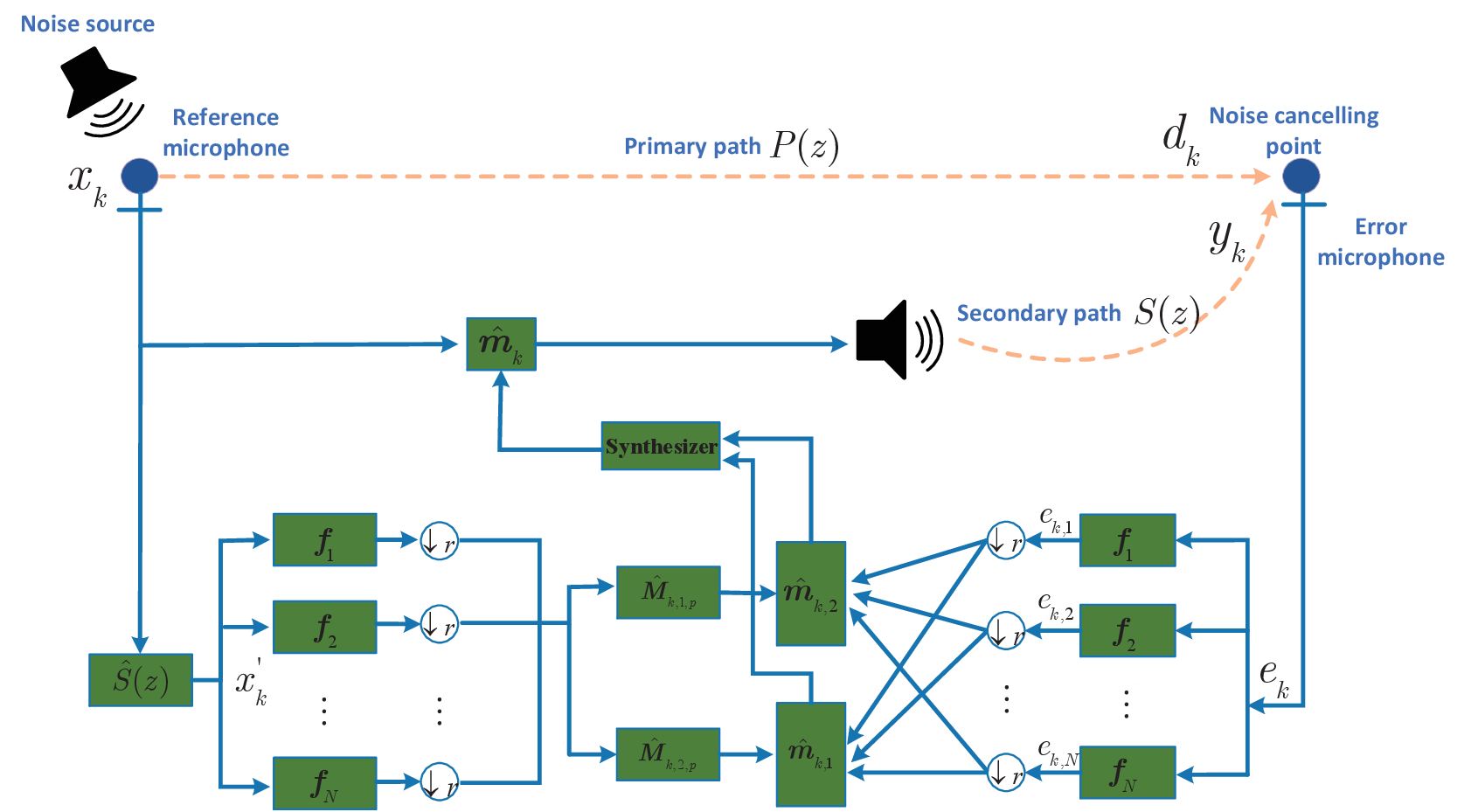}
	\vspace{-1em} \caption{Structure of NKP-FxFoNSPN.} 
	\label{Fig3}
\end{figure}

However, at the noise-cancelling point, only $e_k$ is a measurable signal, while $d_k$ and $y_k$ are unmeasurable. Therefore, unlike the SI model in Fig. 1, we obtain the subband error signals $\{e_{k,j}\}_{j=1}^N$ by passing the error vector $\bm{e}_k=[e_{k}, e_{k-1},..., e_{k-L+1}]^{\text T}$ through the analysis filters $\{\bm{f}_j\}_{j=1}^N$, i.e., 
\vspace{-0.2cm}
\begin{equation}
	\begin{split}
		\begin{array}{rcl}
			\begin{aligned}
				\label{052} 
				\bm{e}_{k,s}\overset{\bigtriangleup}{=}[e_{k,1},e_{k,2},...,e_{k,N}]=\bm{e}_k^{\text T}\bm{F},
			\end{aligned}
		\end{array}
	\end{split}
\end{equation}
where $\bm{e}_{k,s}$ denotes the subband error vector, which is utilized to adjust the weight coefficients of the sub-filters $\hat{\bm m}_{k,1}$ and $\hat{\bm m}_{k,2}$. 

In addition, according to the view proposed by Morgan that the secondary path should be compensated to ensure the stability of the ANC system \cite{1163430}, the input signal $x_k$ should be preprocessed by the secondary path $\hat{S}(z)$ and then used to adjust the weight coefficients of the sub-filters. This is also one of the significant differences from the SI model. Next, we introduce the filtered input matrix ${\bm X}_k^{'}$ of size $D\times L$ as
\begin{equation}
	\begin{split}
		\begin{array}{rcl}
			\begin{aligned}
				\label{053} 
			{\bm X}_k^{'}\overset{\bigtriangleup}{=}\Big[{\bm x}_k^{'}, {\bm x}_{k-1}^{'},..., {\bm x}_{k-L+1}^{'} \Big].
			\end{aligned}
		\end{array}
	\end{split}
\end{equation}

Based on \eqref{053}, the filtered subband input signal can be expressed as
\begin{equation}
	\begin{split}
		\begin{array}{rcl}
			\begin{aligned}
				\label{054} 
				\bm{X}_{k,s}^{'}\overset{\bigtriangleup}{=}\Big[\bm{x}_{k,1}^{'},\bm{x}_{k,2}^{'},...,\bm{x}_{k,N}^{'}\Big]=	{\bm X}_k^{'}\bm{F},
			\end{aligned}
		\end{array}
	\end{split}
\end{equation}
where $\bm{X}_{k,s}^{'}$ denotes the filtered subband input matrix.

To enhance the conciseness of the paper, we directly extend the learning rules of the NKP-FoNSPN and TNKP-FoNSPN algorithms to develop the NKP-FxFoNSPN and TNKP-FxFoNSPN algorithms for ANC applications. The complete pseudo-codes for these algorithms are detailed in Tables 4 and 5.

\textbf{Remark 5}. Similar to the proposed NKP-FoNSPN algorithm. When $\beta=1$, the proposed NKP-FxFoNSPN algorithm degenerates into a novel NKP-FxNSPN algorithm, while its non-NKP decomposition variant becomes a novel FxFoNSPN algorithm. For the weight update formulas of the NKP-FxNSPN and FxFoNSPN algorithms, please refer to \eqref{043}, \eqref{044}, and \eqref{045}.

\textbf{Remark 6}. To highlight the differences between the proposed algorithms and the existing NSPN algorithm, we have provided Table 6, which summarizes the key characteristics of our proposed algorithms. It is worth noting that the symbol "$\bm\times$" does not mean complete failure but rather poor performance. Overall, the fractional-order gradient descent method will enhance the robustness of the algorithm the $0<\alpha\leq 1$ scenario. The NKP technique will improve the overall performance of the algorithm in low-rank systems.

\begin{minipage}[htp]{0.5\textwidth}
	\renewcommand\arraystretch{1.3}
	\scriptsize
	\centering
	\captionof{table}{Pseudo-code of the NKP-FxFoNSPN algorithm}
	\begin{tabular}{lc}
		\hline
		\text{Initialization:} $\hat{\bm m}_{0,1,q}=[\iota\;0\;...\;0]^{\text T},\;q=1,2,...,Q$\\
		\;\;\;\;\;\;\;\;\;\;\;\;\;\;\;\;\;\;\;\;\;$\hat{\bm m}_{0,2,q}=[\iota\;0\;...\;0]^{\text T},\;q=1,2,...,Q$\\
		\hline
		\text{For} each time instant $k$ do:\\
		\;\;\;$	{\bm X}_k^{'}\overset{\bigtriangleup}{=}\Big[{\bm x}_k^{'}, {\bm x}_{k-1}^{'},..., {\bm x}_{k-L+1}^{'} \Big]$\\
		\;\;\;\;\;\;\;\;if $\text{mod}(k,r)==0$\\
		\;\;\;\;\;\;\;\;\;\;\;\;\;\;\;$\bm{X}_{k,s}^{'}\overset{\bigtriangleup}{=}\Big[\bm{x}_{k,1}^{'},\bm{x}_{k,2}^{'},...,\bm{x}_{k,N}^{'}\Big]=	{\bm X}_k^{'}\bm{F}$\\
		\;\;\;\;\;\;\;\;\;\;\;\;\;\;\;$\bm{e}_{k,s}\overset{\bigtriangleup}{=}[e_{k,1},e_{k,2},...,e_{k,N}]=\bm{e}_k^{\text T}\bm{F}$\\
		\;\;\;\;\;\;\;\;\;\;\;\;\;\;\;\;\;\;\;\;\;\;\text{For each subband} $j$\\
		\;\;\;\;\;\;\;\;\;\;\;\;\;\;\;\;\;\;\;\;\;\;\;\;\;\;\;\;$\bm{x}_{k,j,2,q}=\big[\hat{\bm m}_{k,2,q}\otimes\bm{ I}_{D_1}\big]^{\text T}\bm{x}_{k,j}^{'}$\\
		\;\;\;\;\;\;\;\;\;\;\;\;\;\;\;\;\;\;\;\;\;\;\;\;\;\;\;\;$\bm{x}_{k,j,2}=\big[\bm{x}_{k,j,2,1}^{\text T},\bm{x}_{k,j,2,2}^{\text T},...,\bm{x}_{k,j,2,Q}^{\text T}\big]^{\text T}$\\
		\;\;\;\;\;\;\;\;\;\;\;\;\;\;\;\;\;\;\;\;\;\;\;\;\;\;\;\;$\bm{e}_{k,j,1}=\bm{e}_{k,j}$\\
		\;\;\;\;\;\;\;\;\;\;\;\;\;\;\;\;\;\;\;\;\;\;\text{end}\\
		\;\;\;\;\;\;\;\;\;\;\;\;\;\;\;\;\;\;\;\;\;\;$\hat{\bm m}_{k+r,1}=\hat{\bm m}_{k,1}-\mu \sum_{j=1}^N\frac{g(e_{k,j,1})\text{Re}\big\{(-\bm{x}_{k,j,2})^{\beta}\big\}}{\lvert\lvert \bm{x}_{k,j,2}\lvert\lvert_p^p}$\\
		\;\;\;\;\;\;\;\;\;\;\;\;\;\;\;\;\;\;\;\;\;\;\text{For each subband} $j$\\
		\;\;\;\;\;\;\;\;\;\;\;\;\;\;\;\;\;\;\;\;\;\;\;\;\;\;\;\;$\bm{x}_{k,j,1,q}=\big[\bm{ I}_{D_2}\otimes \hat{\bm m}_{k,1,q}\big]^{\text T}\bm{x}_{k,j}^{'}$\\
		\;\;\;\;\;\;\;\;\;\;\;\;\;\;\;\;\;\;\;\;\;\;\;\;\;\;\;\;$\bm{x}_{k,j,1}=\big[\bm{x}_{k,j,1,1}^{\text T},\bm{x}_{k,j,1,2}^{\text T},...,\bm{x}_{k,j,1,Q}^{\text T}\big]^{\text T}$\\
		\;\;\;\;\;\;\;\;\;\;\;\;\;\;\;\;\;\;\;\;\;\;\;\;\;\;\;\;$\bm{e}_{k,j,2}=\bm{e}_{k,j}$\\
		\;\;\;\;\;\;\;\;\;\;\;\;\;\;\;\;\;\;\;\;\;\;\text{end}\\
		\;\;\;\;\;\;\;\;\;\;\;\;\;\;\;\;\;\;\;\;\;\;$\hat{\bm m}_{k+r,2}=\hat{\bm m}_{k,2}-\mu \sum_{j=1}^N\frac{g(e_{k,j,2})\text{Re}\big\{(-\bm{x}_{k,j,1})^{\beta}\big\}}{\lvert\lvert \bm{x}_{k,j,1}\lvert\lvert_p^p}$\\
		\;\;\;\;\;\;\;\;\;\;\;\;\;\;\;\;\;\;\;\;\;\;$\hat{\bm m}_k=\sum_{q=1}^Q\hat{\bm m}_{k,2,q}\otimes\hat{\bm m}_{k,1,q}$\\
		\;\;\;\;\;\;\;\;$\text{else}$\\
		\;\;\;\;\;\;\;\;\;\;\;\;\;\;\;$\hat{\bm m}_{k+r}=\hat{\bm m}_{k}$\\
		\;\;\;\;\;\;\;\;end\\
		end\\
		\hline
	\end{tabular}
\end{minipage}
\begin{minipage}[c]{0.5\textwidth}
	\renewcommand\arraystretch{0.9}
	\scriptsize
	\centering
	\captionof{table}{Pseudo-code of the TNKP-FxFoNSPN algorithm}
	\begin{tabular}{lc}
		\hline
		\text{Initialization:} $\hat{\bm m}_{0,1,q}=[\iota\;0\;...\;0]^{\text T},\;q=1,2,...,Q$\\
		\;\;\;\;\;\;\;\;\;\;\;\;\;\;\;\;\;\;\;\;\;$\hat{\bm m}_{0,2,q}=[\iota\;0\;...\;0]^{\text T},\;q=1,2,...,Q$\\
		\;\;\;\;\;\;\;\;\;\;\;\;\;\;\;\;\;\;\;\;\;$\text{flag=0}$,\;$\text{Flag=0}$ \\
		\hline
		\text{For} each time instant $k$ do:\\
		\;\;\;$	{\bm X}_k^{'}\overset{\bigtriangleup}{=}\Big[{\bm x}_k^{'}, {\bm x}_{k-1}^{'},..., {\bm x}_{k-L+1}^{'} \Big]$\\
		\;\;\;\;\;\;\;\;if $\text{mod}(k,r)==0$\\
		\;\;\;\;\;\;\;\;\;\;\;\;\;\;\;$\bm{X}_{k,s}^{'}\overset{\bigtriangleup}{=}\Big[\bm{x}_{k,1}^{'},\bm{x}_{k,2}^{'},...,\bm{x}_{k,N}^{'}\Big]=	{\bm X}_k^{'}\bm{F}$\\
		\;\;\;\;\;\;\;\;\;\;\;\;\;\;\;$\bm{e}_{k,s}\overset{\bigtriangleup}{=}[e_{k,1},e_{k,2},...,e_{k,N}]=\bm{e}_k^{\text T}\bm{F}$\\
		\;\;\;\;\;\;\;\;\;\;\;\;if $\text{flag}==0$\\
		\;\;\;\;\;\;\;\;\;\;\;\;\;\;\;\;\;\;\;\;\;\;\text{For each subband} $j$\\
		\;\;\;\;\;\;\;\;\;\;\;\;\;\;\;\;\;\;\;\;\;\;\;\;\;\;\;\;$\bm{x}_{k,j,2,q}=\big[\hat{\bm m}_{k,2,q}\otimes\bm{ I}_{D_1}\big]^{\text T}\bm{x}_{k,j}^{'}$\\
		\;\;\;\;\;\;\;\;\;\;\;\;\;\;\;\;\;\;\;\;\;\;\;\;\;\;\;\;$\bm{x}_{k,j,2}=\big[\bm{x}_{k,j,2,1}^{\text T},\bm{x}_{k,j,2,2}^{\text T},...,\bm{x}_{k,j,2,Q}^{\text T}\big]^{\text T}$\\
		\;\;\;\;\;\;\;\;\;\;\;\;\;\;\;\;\;\;\;\;\;\;\;\;\;\;\;\;$\bm{e}_{k,j,1}=\bm{e}_{k,j}$\\
		\;\;\;\;\;\;\;\;\;\;\;\;\;\;\;\;\;\;\;\;\;\;\text{end}\\
		\;\;\;\;\;\;\;\;\;\;\;\;\;\;\;\;\;\;\;\;\;\;$\hat{\bm m}_{k+r,1}=\hat{\bm m}_{k,1}-\mu \sum_{j=1}^N\frac{g(e_{k,j,1})\text{Re}\big\{(-\bm{x}_{k,j,2})^{\beta}\big\}}{\lvert\lvert \bm{x}_{k,j,2}\lvert\lvert_p^p}$\\
		\;\;\;\;\;\;\;\;\;\;\;\;\;\;\;\;\;\;\;\;\;\;\text{For each subband} $j$\\
		\;\;\;\;\;\;\;\;\;\;\;\;\;\;\;\;\;\;\;\;\;\;\;\;\;\;\;\;$\bm{x}_{k,j,1,q}=\big[\bm{ I}_{D_2}\otimes \hat{\bm m}_{k,1,q}\big]^{\text T}\bm{x}_{k,j}^{'}$\\
		\;\;\;\;\;\;\;\;\;\;\;\;\;\;\;\;\;\;\;\;\;\;\;\;\;\;\;\;$\bm{x}_{k,j,1}=\big[\bm{x}_{k,j,1,1}^{\text T},\bm{x}_{k,j,1,2}^{\text T},...,\bm{x}_{k,j,1,Q}^{\text T}\big]^{\text T}$\\
		\;\;\;\;\;\;\;\;\;\;\;\;\;\;\;\;\;\;\;\;\;\;\;\;\;\;\;\;$\bm{e}_{k,j,2}=\bm{e}_{k,j}$\\
		\;\;\;\;\;\;\;\;\;\;\;\;\;\;\;\;\;\;\;\;\;\;\text{end}\\
		\;\;\;\;\;\;\;\;\;\;\;\;\;\;\;\;\;\;\;\;\;\;$\hat{\bm m}_{k+r,2}=\hat{\bm m}_{k,2}-\mu \sum_{j=1}^N\frac{g(e_{k,j,2})\text{Re}\big\{(-\bm{x}_{k,j,1})^{\beta}\big\}}{\lvert\lvert \bm{x}_{k,j,1}\lvert\lvert_p^p}$\\
		\;\;\;\;\;\;\;\;\;\;\;\;\;\;\;\;\;\;\;\;\;\;$\hat{\bm m}_k=\sum_{q=1}^Q\hat{\bm m}_{k,2,q}\otimes\hat{\bm m}_{k,1,q}$\\
		\;\;\;\;\;\;\;\;\;\;\;\; $\text{else}$\\
		\;\;\;\;\;\;\;\;\;\;\;\;\;\;\;\;\;\;\;$\hat{\bm m}_{k+r}=\hat{\bm m}_{k}-\mu_b \sum_{j=1}^N \frac{\lvert e_{k,j}\lvert^{p-2}e_{k,j}\text{Re}\big\{(-\bm{x}_{k,j}^{'})^{\beta}\big\}}{\lvert\lvert \bm{x}_{k,j}^{'}\lvert\lvert_p^p}$\\
		\;\;\;\;\;\;\;\;\;\;\;\; $\text{end}$\\
		\;\;\;\;\;\;\;\;else\\
		\;\;\;\;\;\;\;\;\;\;\;\;\;\;\;$\hat{\bm m}_{k+r}=\hat{\bm m}_{k}$\\
		\;\;\;\;\;\;\;\;end\\
		\;\;\;\;\;\;\;\;\text{if} $\text{Flag}=0$\\
		\;\;\;\;\;\;\;\;\;\;\;\text{if} $\text{ANR}_k \leq \rho$\\
		\;\;\;\;\;\;\;\;\;\;\;\;\;\;\;$\text{flag}=1$;\\
		\;\;\;\;\;\;\;\;\;\;\;\;\;\;\;$\text{Flag}=1$;\\
		\;\;\;\;\;\;\;\;\;\;\;\text{else}\\
		\;\;\;\;\;\;\;\;\;\;\;\;\;\;\;$\text{flag}=0$;\\
		\;\;\;\;\;\;\;\;\;\;\;end\\
		\;\;\;\;\;\;\;\;end\\
		end\\
		\hline
	\end{tabular}
\end{minipage}
\begin{table*}[htp]
	\renewcommand\arraystretch{0.8}
	\tabcolsep = 0.3cm
	\setlength{\abovecaptionskip}{0cm}
	\setlength{\belowcaptionskip}{0cm}
	\begin{center}
		\caption{Comparison of proposed and existing methods.}
		\begin{tabular}{c c c c c c c} 
			\hline
			\rowcolor{gray!30}\footnotesize{Algorithms} &$0<\alpha\leq1$ & $1<\alpha<2$ & AEC & ANC & Low-rank system & High-rank system   \\			
			\hline
			\footnotesize{NSPN} &$\bm\times$ &$\checkmark$&$\checkmark$&$\checkmark$ &$\times$&$\checkmark$\\
			\rowcolor{gray!30}\footnotesize{FoNSPN} &$\checkmark$ &$\checkmark$&$\checkmark$&$\checkmark$ &$\bm\times$&$\checkmark$\\
			\footnotesize{NKP-NSPN} &$\bm\times$ &$\checkmark$&$\checkmark$&$\bm\times$ &$\checkmark$&$\bm\times$\\
			\rowcolor{gray!30}\footnotesize{NKP-FoNSPN} &$\checkmark$ &$\checkmark$&$\checkmark$&$\bm\times$ &$\checkmark$&$\bm\times$\\
			\footnotesize{TNKP-FoNSPN} &$\checkmark$ &$\checkmark$&$\bm\times$&$\checkmark$ &$\checkmark$&$\bm\times$\\		
			\hline
		\end{tabular}
	\end{center}
\end{table*}
\section{Simulation Result}
In this section, we will substantiate the efficacy of the proposed algorithms through extensive simulations. Based on the cosine-modulated principle, the analysis filters $\{\bm{f}_j\}_{j=1}^N$ with subband number $N=4$ and length $L=33$ are designed using the prototype filter. In addition, the length of the adaptive filter $\hat{\bm m}_k$ is $D=500$, and the lengths of the sub-filters $\hat{\bm m}_{k,1}$ and $\hat{\bm m}_{k,2}$ are $D_1=25$ and $D_2=20$, respectively.
\begin{figure}
	\centering  
	\includegraphics[scale=0.42] {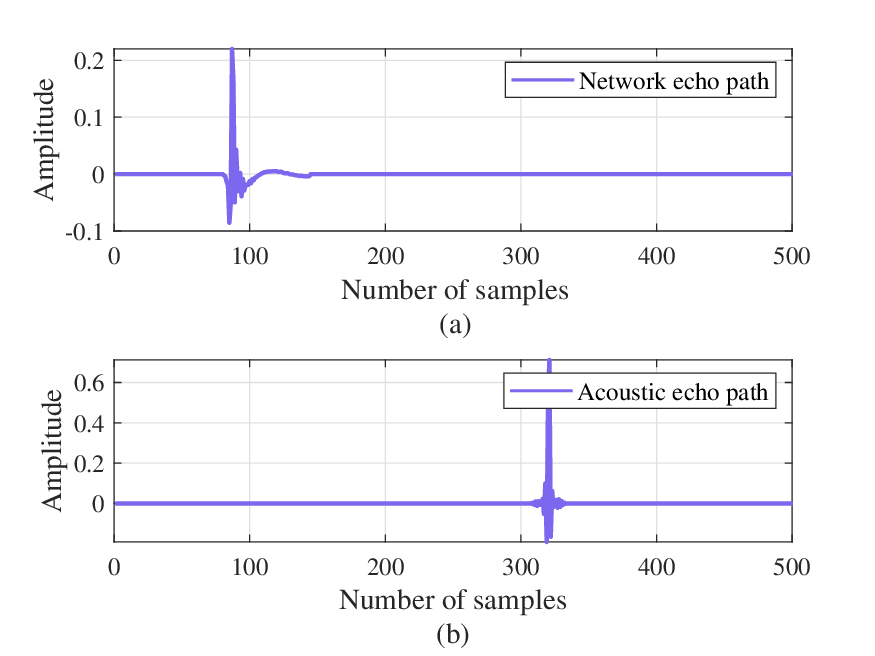}  								  
	\caption{IR of the unknown system $\bm{m}_0$ with $D=500$. (a) network echo path is chosen from G.168 Recommendation \cite{ITU} and (b) acoustic echo path \cite{6802370}.}
	\label{Fig4}
\end{figure}
\subsection{System Identification}
Fig. 5 depicts the IR of the unknown system $\bm{m}_{0}$ of length $D=500$, where Fig. 5 (a) denotes the acoustic echo path and (b) represents the network echo path. Unless otherwise specified, the input signal is generated by passing zero-mean Gaussian noise through a first-order autoregressive model with a pole at 0.9. The additive noise $v_k$ can be modeled as a symmetric $\alpha$-stable random process. In addition, we use the normalized mean-square deviation (NMSD) to evaluate the performance of the algorithms, which is defined as
\begin{equation}
	\begin{split}
		\begin{array}{rcl}
			\begin{aligned}
				\label{056}
				\ \text{NMSD}(\text{dB})\overset{\bigtriangleup}{=}20\text{log}_{10}\text{E}\Big\{\frac{\lvert\lvert \bm{m}_0-\hat{\bm m}_k\lvert\lvert_2}{\lvert\lvert \bm{m}_0\lvert\lvert_2}\Big\}.
			\end{aligned}
		\end{array}
	\end{split}
\end{equation}
The results are averaged over 50 Monte Carlo trials.
\begin{figure}
	\centering  
	\includegraphics[scale=0.42] {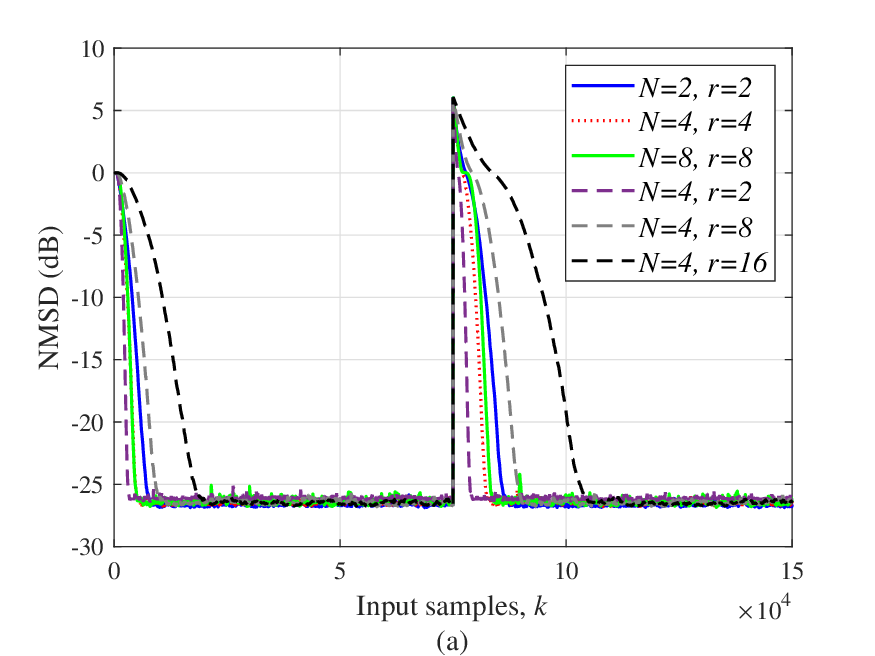}
	\hspace{0.1ex}									 
	\includegraphics[scale=0.42] {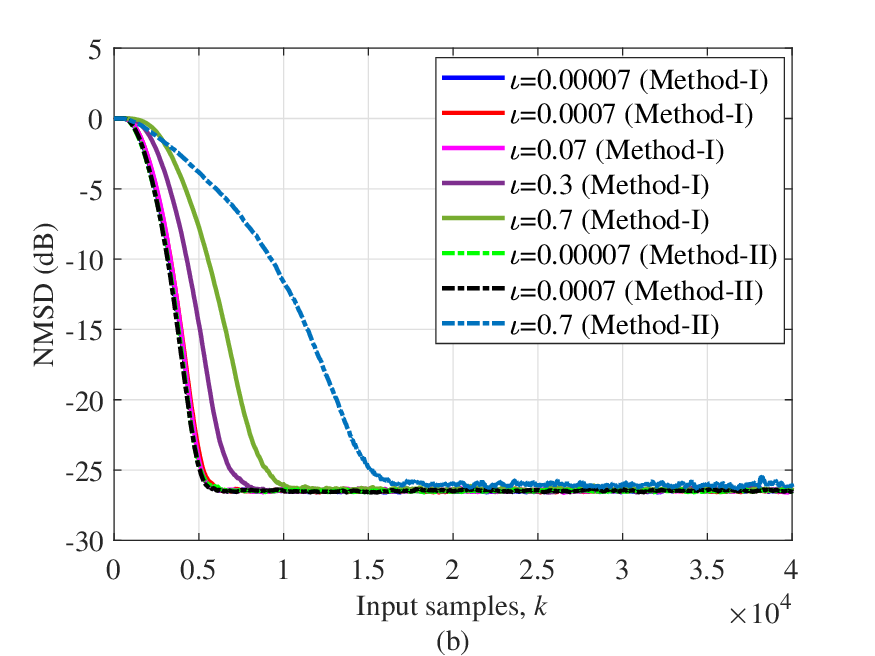}	  								  
	\caption{NMSD curves of the NKP-FoNSPN algorithm for different $N$, $r$, and $\iota$. $\alpha=1.5$ and $\gamma=1/60$ for the additive noise; $\mu=0.01$, $p=1.4$, $\beta=1.1$, and $Q=2$ for the NKP-FoNSPN algorithm.}
	\label{Fig5}
\end{figure}

$A.$ {\it Parameter Selection } 

Fig. 6(a) examines the influence of parameters $N$ and $r$ on the NMSD performance of the NKP-FoNSPN algorithm in acoustic echo path identification. Increasing the number of subbands $N$ enhances the decorrelation capability for correlated inputs and accelerates convergence. However, the convergence performances of the NKP-FoNSPN algorithms with $N = 4$ and $N = 8$ are nearly identical. Therefore, $N=4$ is selected for all subsequent simulations. Furthermore, under the condition of $N = 4$, the NKP-FoNSPN algorithm converges faster as $r$ decreases; however, its computational complexity also increases. Once $r$ is smaller than $N$, the improvement in convergence rate becomes marginal; hence, $r = 4$ is adopted. In addition, we verify the tracking capability of the algorithm when the IR of the unknown system in the middle of the input sample changes from $\bm{m}_0$ to $-\bm{m}_0$. The results demonstrate that the NKP-FoNSPN algorithm exhibits good tracking performance when faced with a rapidly changing unknown system. Fig. 6(b) examines the influence of initial value $\iota$ on the NMSD performance of the NKP-FoNSPN algorithm with Method-I and Method-II. Clearly, different initial values $\iota$ will affect the convergence performance of the NKP-FoNSPN algorithm. Among them, the NKP-FoNSPN algorithms with $\iota= 0.00007$, 0.0007, and 0.07 have a slightly faster convergence rate compared to $\iota= 0.3$ and 0.7. In addition, the NMSD performance of the NKP-FoNSPN algorithm with $\iota= 0.00007$ (Method-I), $\iota= 0.0007$ (Method-I), $\iota= 0.00007$ (Method-II), and $\iota= 0.0007$ (Method-II) is completely identical. For the subsequent simulations, we set $\iota$ equal to 0.0007.

Fig. 7 examines the influence of different fractional-order parameters $\beta$ on the NMSD performance of the NKP-FoNSPN algorithm for network echo path identification. Based on the experimental parameters and \eqref{050},  the fractional-order $\beta$ that ensures stable convergence of the proposed NKP-FoNSPN algorithm must lie within $(0.65, 1.4]$. Both $\beta = 0.8$ and $\beta = 1.1$ fall within this range, enabling stable convergence. In contrast, $\beta = 0.3$ and $\beta = 1.8$ lie outside these bounds, leading to unstable convergence or divergence.
\begin{figure}
	\centering  
	\includegraphics[scale=0.42] {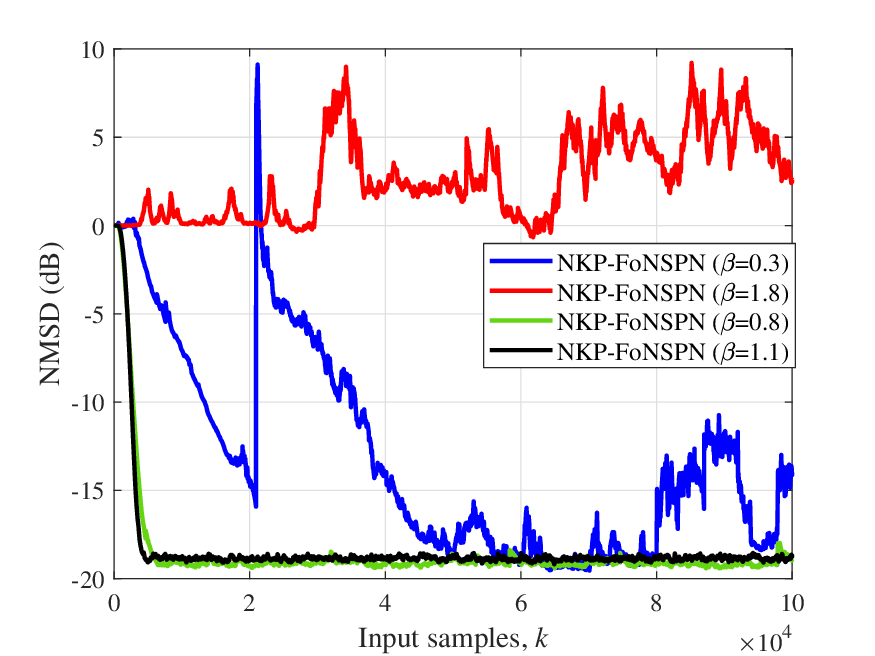}  								  
	\caption{NMSD curves of the NKP-FoNSPN algorithm for different $\beta$. Refering to Fig. 6 for other parameters of the algorithms.}
	\label{Fig6}
\end{figure}

$B.$ {\it Transformation NKP decomposition technique}

Fig. 8 examines the efficacy of the TNKP decomposition technique in NKP-NLMS and NKP-FoNSPN algorithms for acoustic echo path identification. As shown in Fig. 8 (a), the NKP-NLMS algorithm ($\mu=0.08$) converges faster than NLMS ($\mu=1$) while maintaining identical steady-state misadjustment. Note that $\mu=1$ yields the fastest convergence for the NLMS algorithm. However, step-size adjustment alone cannot enable the NKP-NLMS algorithm to achieve the steady-state misadjustment level of the NLMS algorithm with $\mu=0.3$. The TNKP decomposition technique addresses this limitation, yielding the TNKP-NLMS algorithm that combines NKP's rapid convergence with NLMS's low steady-state misadjustment at small step-sizes. As shown in Fig. 8 (b), the NKP-FoNSPN algorithm ($\mu=0.005$) achieves the same steady-state misadjustment as the FoNSPN algorithm ($\mu=0.028$) but significantly improves the convergence rate. Although NKP-FoNSPN shares the fundamental limitations of NKP-NLMS, its enhanced variant TNKP-FoNSPN achieves superior convergence rate and lower steady-state misadjustment relative to benchmark algorithms.
\begin{figure}
	\centering  
	\includegraphics[scale=0.42] {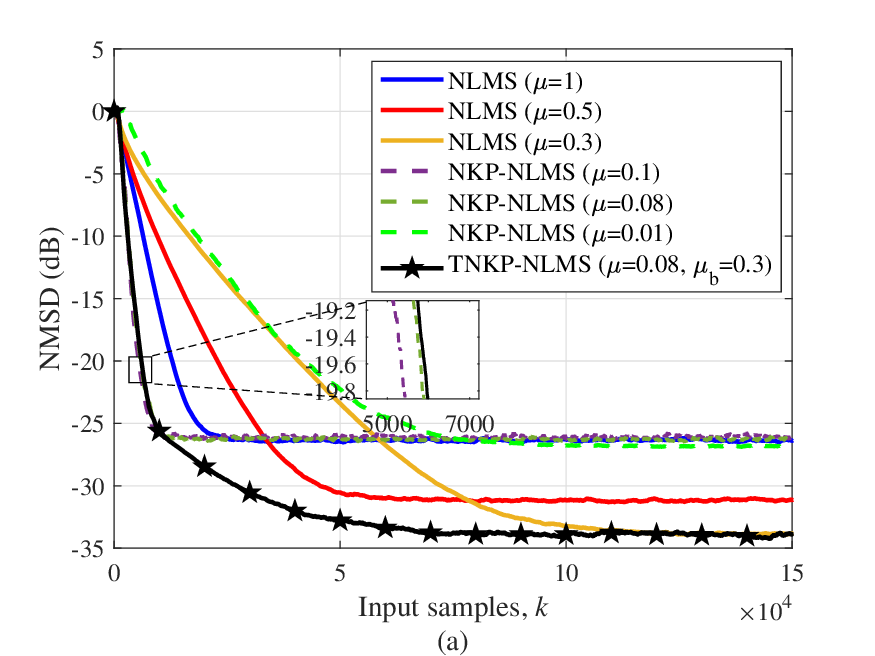}  
	\hspace{0.1ex}									 
	\includegraphics[scale=0.42] {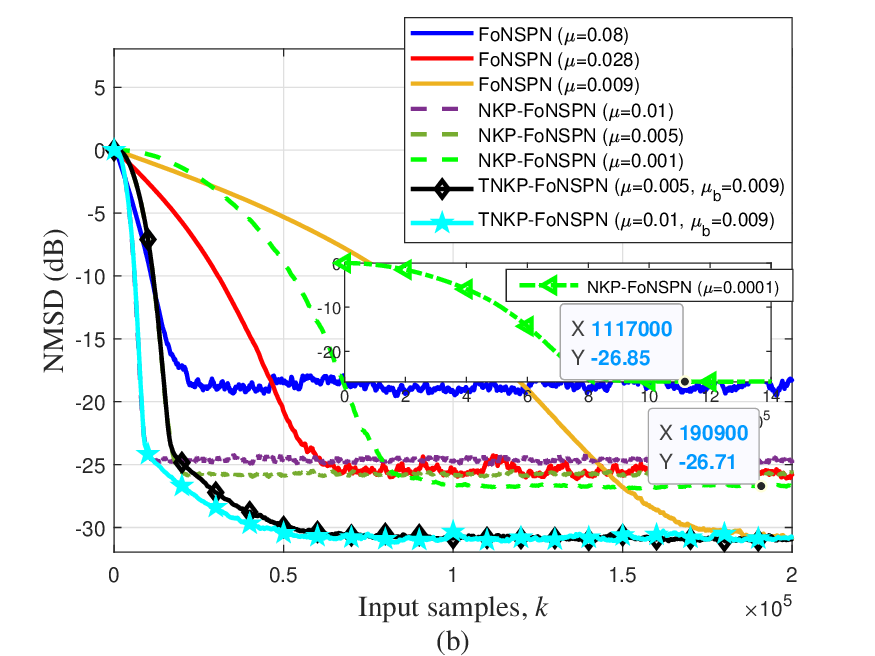}								  
	\caption{Testing the effectiveness of the transformation NKP decomposition technique. (a) The TNKP-NLMS algorithm and (b) the TNKP-FoNSPN algorithm. $\alpha=2$, $\gamma=1/60$, and $Q=2$ for Fig. 8 (a); $\alpha=0.75$, $\gamma=1/60$, $p=0.7$, $Q=2$, and $\beta=0.65$ for Fig. 8 (b). }
	\label{Fig7}
\end{figure}
\begin{figure}
	\centering  
	\includegraphics[scale=0.42] {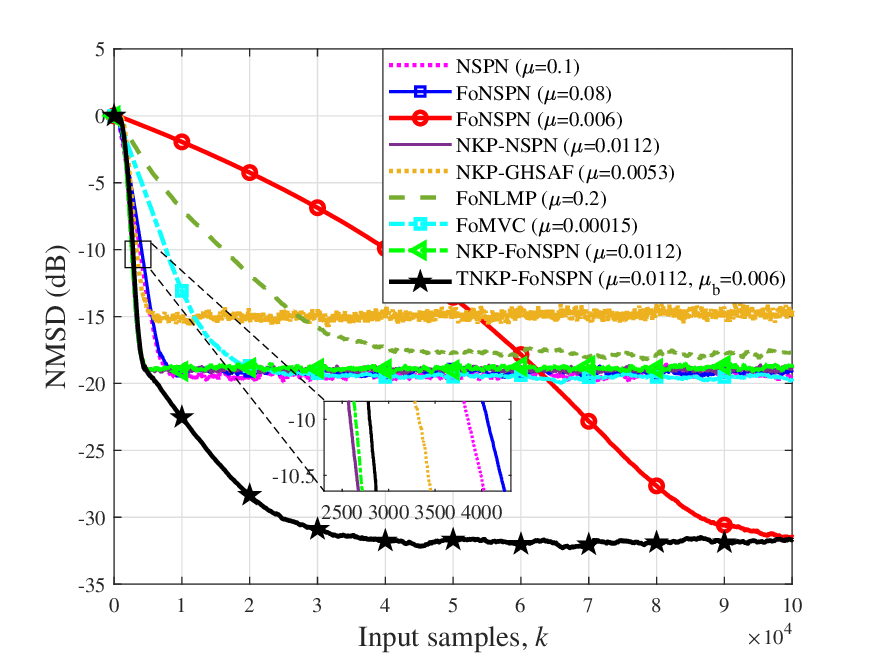}  								  
	\caption{Comparison of NMSD learning curves for $\alpha$-stable noise with $\alpha=1.5$ and $\gamma=1/60$. $Q=2$ for NKP decomposition; $p=1.4$ for the NSPN algorithm; $p=1.4$ and $\beta=1.1$ for the FoNSPN, FoNLMP, NKP-FoNSPN, and TNKP-FoNSPN algorithms; $\hat{\rho}=0.6$, $\alpha^{'}=1.4$, and $\tau=0.2$ for the FoMVC algorithm; $\lambda=0.99$ and $\alpha^{'}=2$ for the NKP-GHSAF algorithm. }
	\label{Fig8}
\end{figure}

$C.$ {\it Network Echo Path} 

Fig. 9 compares the NMSD learning curves for the NSPN \cite{11113}, FoNSPN, NKP-NSPN, NKP-GHSAF \cite{9170797}, FoNLMP \cite{3}, FoMVC \cite{8}, NKP-FoNSPN, and TNKP-FoNSPN algorithms. The FoMVC algorithm achieves significantly faster convergence than the FoNLMP algorithm. At $\mu=0.08$, the FoNSPN algorithm matches the FoMVC algorithm's steady-state misadjustment while converging faster due to its subband-based architecture. Notably, for $1<\alpha \leq 2$ scenarios, the NSPN and FoNSPN ($\mu=0.08$) algorithms exhibit nearly identical performance in terms of convergence rate and steady-state misadjustment. However, when $0<\alpha \leq 1$, the performance of the NSPN algorithm deteriorates sharply, which is verified by the simulation results in Fig. 10. Additionally, the NKP-NSPN algorithm (equivalent to NKP-FoNSPN with $\beta$ = 1) demonstrates performance comparable to the NKP-FoNSPN algorithm. Compared to the FoNSPN algorithm with $\mu=0.08$, the proposed NKP-FoNSPN algorithm converges slightly faster. Compared to the NKP-GHSAF algorithm, the proposed NKP-FoNSPN algorithm maintains significantly lower steady-state misadjustment. Although the NKP-FoNSPN algorithm converges substantially faster than the FoNSPN algorithm ($\mu=0.006$), it performs poorly in terms of steady-state NMSD. Crucially, the proposed TNKP-FoNSPN algorithm achieves both the fastest convergence rate and the lowest steady-state misadjustment among all compared algorithms.
\begin{figure}
	\centering  
	\includegraphics[scale=0.42] {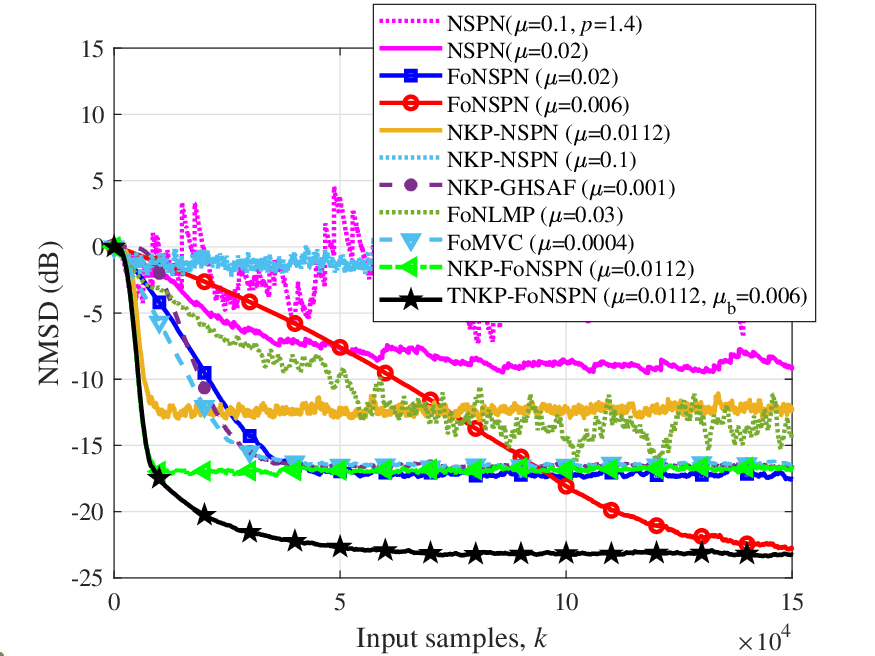}  								  
	\caption{Comparison of NMSD learning curves for $\alpha$-stable noise with $\alpha=0.75$ and $\gamma=1/60$. $\alpha^{'}=1.1$ for the FoMVC algorithm; $p=0.7$ and $\beta=0.65$ for the FoNSPN, FoNLMP, NKP-FoNSPN, and TNKP-FoNSPN algorithms. Refering to Fig. 9 for other parameters of the algorithms.}
	\label{Fig9}
\end{figure}

Furthermore, we evaluate the performance of the algorithms under $\alpha$-stable noise ($\alpha$=0.75). As shown in Fig. 10, the NSPN algorithm ($\mu$=0.1) cannot converge stably, however, when a small step-size is directly applied to the NSPN algorithm (NSPN with $\mu$=0.02), stable convergence is achieved, but the performance of NSPN ($\mu$=0.02) is significantly inferior to that of FoNSPN ($\mu$=0.02) under the same small step-size in terms of convergence rate and steady-state misadjustment. Although the NKP-NSPN algorithm ($\mu$=0.0112) achieves stable convergence compared to the original NSPN algorithm ($\mu$=0.1), this stability is not attributed to the NKP technique itself but rather to the adoption of a small step-size. In addition, the FoNSPN algorithm ($\mu$=0.02) maintains a faster convergence than the FoNLMP algorithm but achieves a slightly slower convergence than the FoMVC algorithm. Unlike in Fig. 9, the NKP-NSPN algorithm exhibits significant performance degradation relative to the NKP-FoNSPN algorithm in this noise environment, as the NSPN algorithm performs poorly in scenarios where $0<\alpha\leq 1$. Additionally, the proposed NKP-FoNSPN algorithm demonstrates superior transient convergence performance compared to the FoMVC, NKP-GHSAF, and FoNSPN algorithms. Crucially, the proposed TNKP-FoNSPN algorithm successfully achieves the target steady-state NMSD of FoNSPN ($\mu$=0.006) while preserving the rapid convergence characteristic of NKP-FoNSPN method.
\begin{figure}
	\centering  
	\includegraphics[scale=0.42] {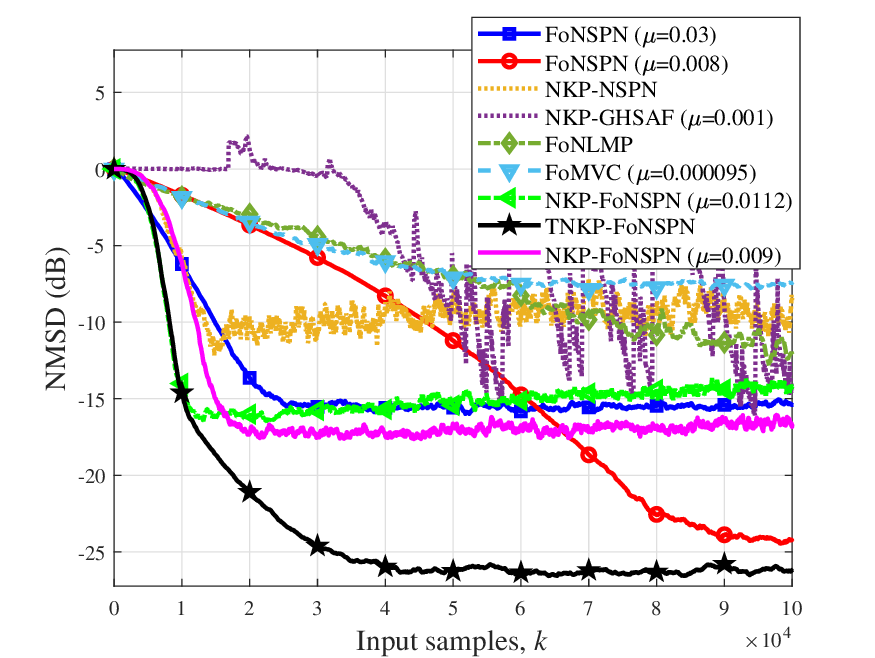}  								  
	\caption{Comparison of NMSD learning curves for Cauchy noise input with $\alpha=1$ and $\gamma=1/10$ and $\alpha$-stable noise with $\alpha=0.75$ and $\gamma=1/60$. $\alpha^{'}=1.1$ for the FoMVC algorithm; $p=0.7$ and $\beta=0.65$ for the FoNSPN, FoNLMP, NKP-FoNSPN, and TNKP-FoNSPN algorithms. Refering to Fig. 10 for other parameters of the algorithms.}
	\label{Fig10}
\end{figure}

In Fig. 11, the input signal is generated by passing Cauchy noise with $\alpha=1$ and $\gamma=1/10$ through a first-order autoregressive model with a pole at 0.9. As shown, the NKP-NSPN and NKP-GHSAF algorithms perform poorly and fail to converge stably under impulsive noise inputs, as outliers in the impulsive noise induce erroneous gradient updates via the input signal. Interestingly, FoSGD-based adaptive filtering algorithms possess good robustness to impulsive noise inputs. Specifically, the FoNSPN algorithm with $\mu=0.03$ achieves faster convergence than both the FoNLMP and FoMVC algorithms. Furthermore, the proposed NKP-FoNSPN ($\mu=0.0112$) algorithm exhibits a faster convergence rate than the FoNSPN algorithm, a phenomenon attributable to its inheritance of the structural advantages inherent in the NKP framework, but the steady-state performance shows a divergent trend. The reason for this phenomenon lies in the inappropriate selection of the step-size. Clearly, the NKP-FoNSPN ($\mu=0.009$) algorithm achieves faster convergence rate and lower steady-state misadjustment compared to competing algorithms. In conclusion, the proposed TNKP-FoNSPN algorithm demonstrably achieves both faster convergence and lower steady-state misadjustment compared to all algorithms.

$D.$ {\it Acoustic Echo Path}

In this subsection, we evaluate the NMSD performance of all algorithms in the acoustic echo path scenario. As shown in Figs. 12 (a) and (b), the results yield conclusions similar to those from the network echo path scenario. Overall, the FoNSPN algorithm achieves faster convergence than the FoNLMP and FoMVC algorithms. Furthermore, the proposed NKP-FoNSPN algorithm not only achieves a faster convergence rate than the FoNSPN algorithm but also exhibits lower steady-state misadjustment than the NKP-NSPN algorithm under Gaussian inputs. Significantly, only the TNKP-FoNSPN algorithm simultaneously achieves both rapid convergence and low steady-state misadjustment. Notably, non-FoSGD algorithms (i.e., NKP-NSPN and NKP-GHSAF) fail to converge stably or even diverge when the input signal is characterized by impulsive noise.
\begin{figure}
	\centering  
	\includegraphics[scale=0.42] {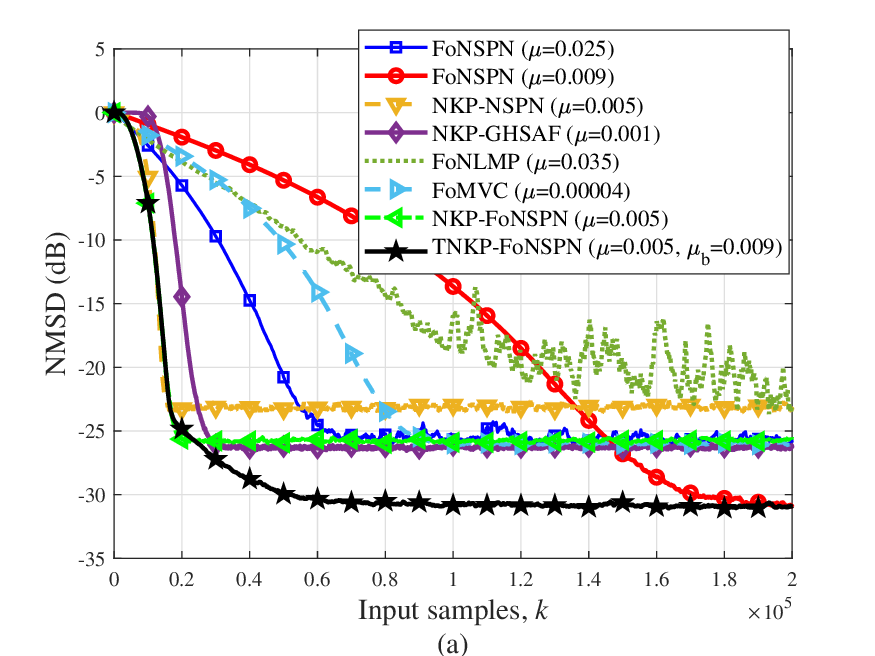} 
	\hspace{0.1ex} 
		\includegraphics[scale=0.42] {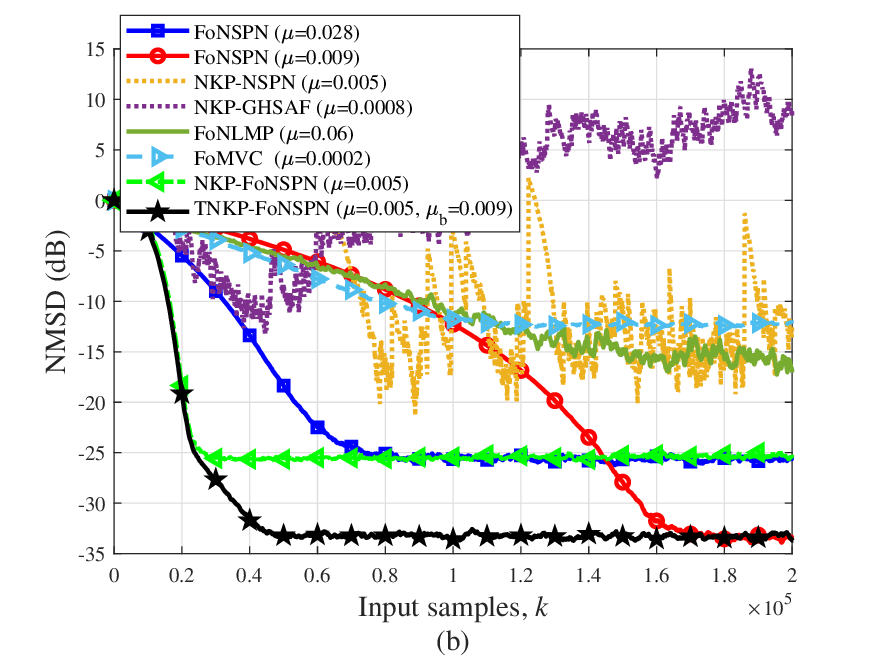} 								  
	\caption{Comparison of NMSD learning curves for $\alpha$-stable noise with $\alpha=0.75$ and $\gamma=1/60$. (a) $\alpha^{'}=1.1$ for the FoMVC algorithm; $\lambda=0.9$ and $\alpha^{'}=0.5$ for the NKP-GHSAF algorithm; $p=0.7$ and $\beta=0.65$ for the FoNSPN, FoNLMP, NKP-FoNSPN, and TNKP-FoNSPN algorithms. Refering to Fig. 9 for other parameters of the algorithms. (b) Cauchy noise input with $\alpha=1$ and $\gamma=1/10$, refering to (a) for other parameters of the algorithms.}
	\label{Fig12}
\end{figure}

$E.$ {\it High-rank system}
\begin{figure}
	\centering  
	\includegraphics[scale=0.42] {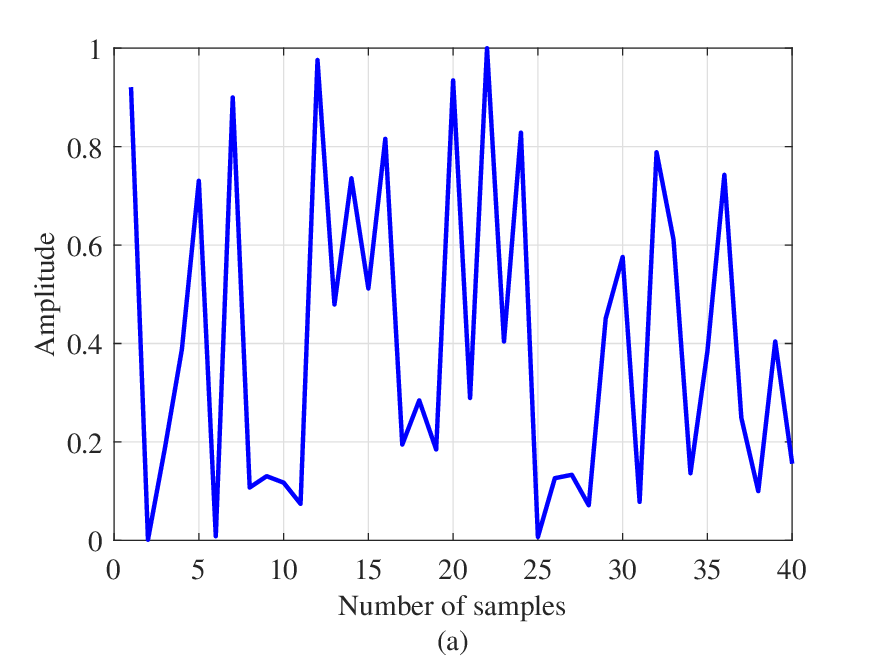}  
	\hspace{0.1ex}									 
	\includegraphics[scale=0.42] {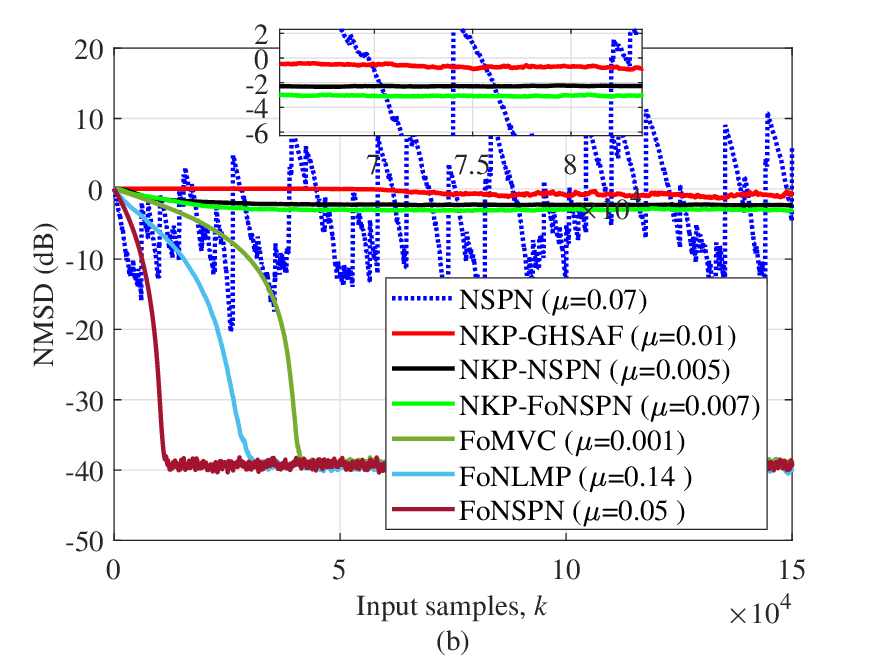}								  
	\caption{(a) IR of the non-low-rank system; (b) Comparison of NMSD learning curves for $\alpha$-stable noise with $\alpha=0.75$ and $\gamma=1/60$. $D_1=8$, $D_2=5$, $Q=4$ for NKP decomposition; $p=1.4$ for the NSPN algorithm; $p=0.7$ and $\beta=0.65$ for the FoNSPN, FoNLMP, and NKP-FoNSPN algorithms; $\hat{\rho}=0.6$, $\alpha^{'}=1.4$, and $\tau=0.2$ for the FoMVC algorithm; $\lambda=0.99$ and $\alpha^{'}=2$ for the NKP-GHSAF algorithm.}
	\label{Fig7}
\end{figure}

The aforementioned simulation experiments validate the NMSD performance of the proposed NKP-NSPN, NKP-FoNSPN, and TNKP-FoNSPN algorithms under sparse low-rank systems. Clearly, the algorithms based on the NKP technique achieve faster convergence rate than non-NKP algorithms, which is consistent with the conclusions reported in relevant literature on NKP-based adaptive filtering algorithms. In this subsection, we further verify the performance of the NKP technique in non-low-rank system. Fig. 13(a) depicts a non-low-rank system, whose weights are randomly generated. Furthermore, as shown in Fig. 13(b), the FoNSPN algorithm exhibits faster convergence rate than the FoMVC and FoNLMP algorithms. However, due to the structural constraints of the NKP technique, the performance of the NKP-FoNSPN, NKP-NSPN, and NKP-GHSAF algorithms deteriorates sharply in non-low-rank system, while the NKP-FoNSPN algorithm still outperforms the NKP-NSPN and NKP-GHSAF algorithms.
\subsection{Acoustic Echo Cancellation}
\begin{figure}
	\centering  
	\includegraphics[scale=1.2] {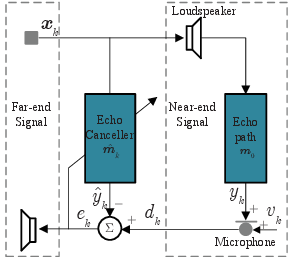}  								  
	\caption{Structure of the adaptive echo canceler.}
	\label{Figb}
\end{figure}
For the echo cancellation system described in Fig. 14, $\bm{x}_k$ represents the speech input signal. By passing $\bm{x}_k$ through the echo path $\bm{m}_0$, we get the echo signal $y_k$. Using the same input signal, the output of the echo canceller with filter $\hat{\bm m}_r$ is $\hat{y}_k$, which is the estimated value of the echo signal $y_k$. Subsequently, by subtracting $\hat{y}_k$ from the desired signal $d_k$, a higher-quality speech signal $e_k$, free from echo interference, is obtained.

In this subsection, we evaluate the performance of the proposed algorithms in AEC scenarios. As shown in Figs. 15 (a) and (b), unlike in the SI scenario, the performance of the TNKP-FoNSPN algorithm in the AEC scenario is inferior to that of the NKP-FoNSPN algorithm. This is because the small step-size FoNSPN ($\mu=0.08$) algorithm exhibits a higher steady-state misadjustment than the NKP-FoNSPN algorithm, leading the TNKP-FoNSPN algorithm to inherit the high steady-state misadjustment of the FoNSPN algorithm during the steady-state phase. Therefore, the TNKP technique is not applicable here. Notably, the results yield conclusions similar to those from the SI scenario. Overall, the proposed NKP-FoNSPN algorithm achieves the fastest convergence rate and the lowest steady-state misadjustment compared to the competing algorithms.
\begin{figure}
	\centering  
	\includegraphics[scale=0.42] {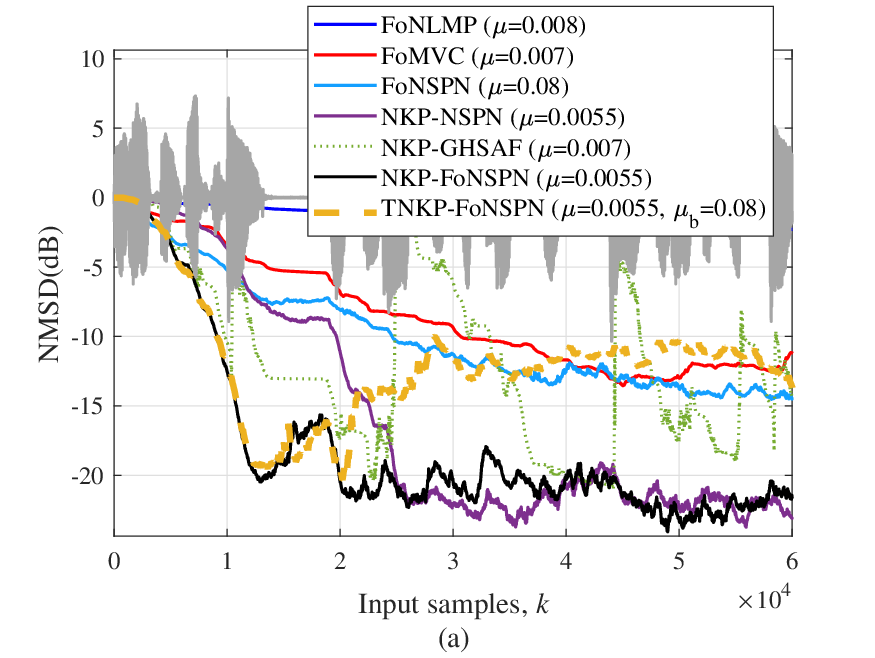}  
	\hspace{0.1ex}									 
	\includegraphics[scale=0.42] {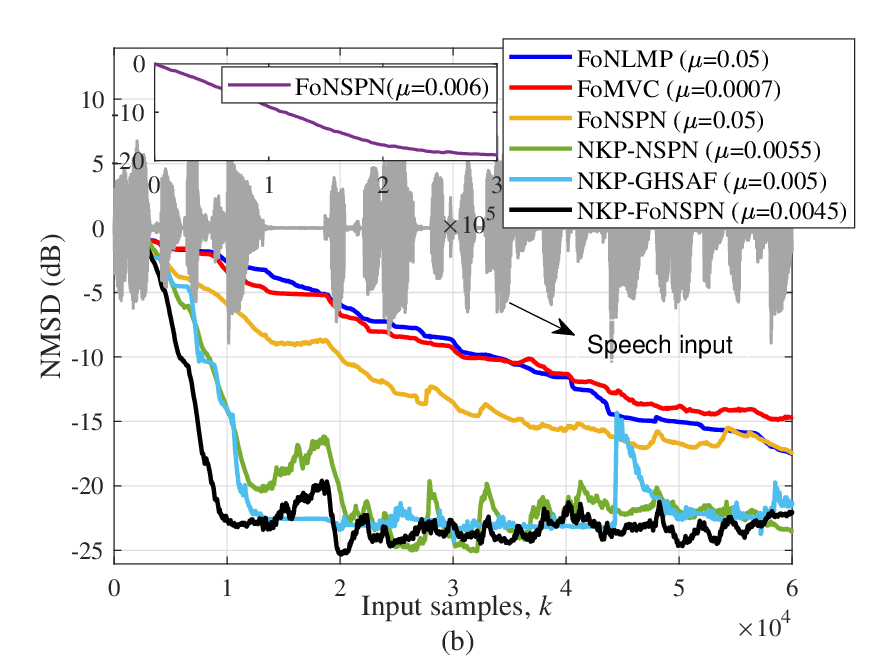}								  
	\caption{NMSD learning curves of different algorithm in AEC scenarios. (a) $\alpha$-stable noise with $\alpha=0.75$ and $\gamma=1/60$, $p=1$ for the NKP-NSPN algorithm, $\rho=0.75$ for the FoMVC algorithm, and refering to Fig. 10 for other parameters of the algorithms. (b) $\alpha$-stable noise with $\alpha=1.5$ and $\gamma=1/60$, $\rho=0.75$ for the FoMVC algorithm, and refering to Fig. 9 for parameters of the algorithms. }
	\label{Figa}
\end{figure}

\subsection{Active Noise Control}
In this subsection, we evaluate the noise control capability of the proposed algorithms using a large number of real noises (e.g., helicopter noise, pile driver noise, gunshots, and traction substation noise). The primary parameter settings for the algorithms are provided in Table 7. Fig. 16 depicts the time-domain IRs of the primary and secondary paths. Additionally, the algorithm's performance indicator average noise reduction (ANR) is defined as \cite{5535133}:
\begin{equation}
	\begin{split}
		\begin{array}{rcl}
			\begin{aligned}
				\label{057}
				\ \text{ANR}(k)\overset{\bigtriangleup}{=}20\text{log}\Big\{\frac{Z_e(k)}{Z_d(k)} \Big\},
			\end{aligned}
		\end{array}
	\end{split}
\end{equation}
where $ Z_e(k)=\eta Z_e(k-1) + (1-\eta)\lvert e_k\lvert$ and $Z_d(k)=\eta Z_d(k-1) + (1-\eta)\lvert d_k\lvert$  denote the averaged magnitude of residual error and desired signal, and $\eta=0.999$.
\begin{figure}
	\centering  
	\includegraphics[scale=0.42] {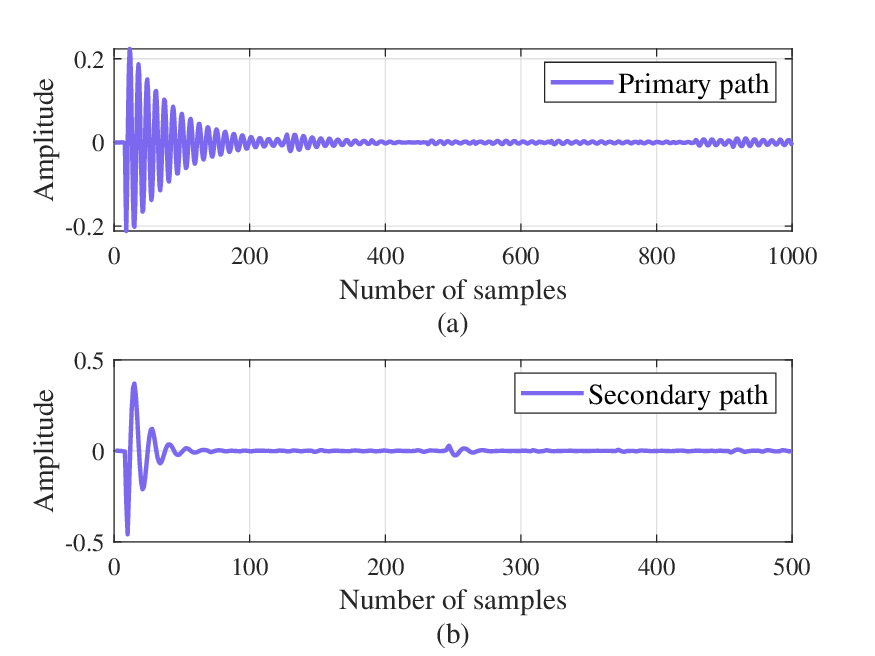}  								  
	\caption{Time-domain IRs of ANC system \cite{wang2025family}. (a) Primary path and (b) secondary path.}
	\label{Fig13}
\end{figure}

$A.$ {\it Pink Noise}
\begin{figure}
	\centering  
	\includegraphics[scale=0.42] {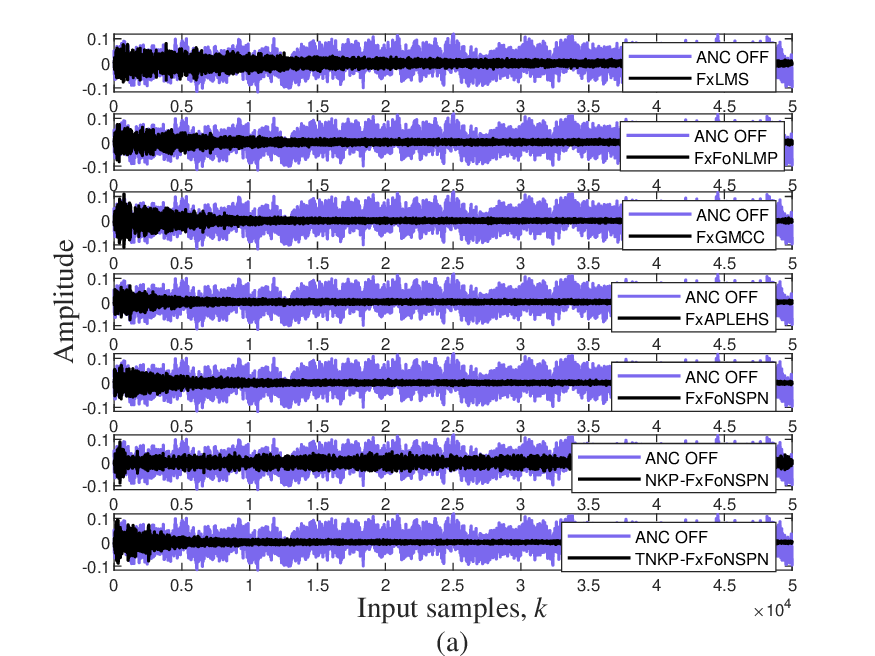}  
	\hspace{0.1ex}									 
	\includegraphics[scale=0.42] {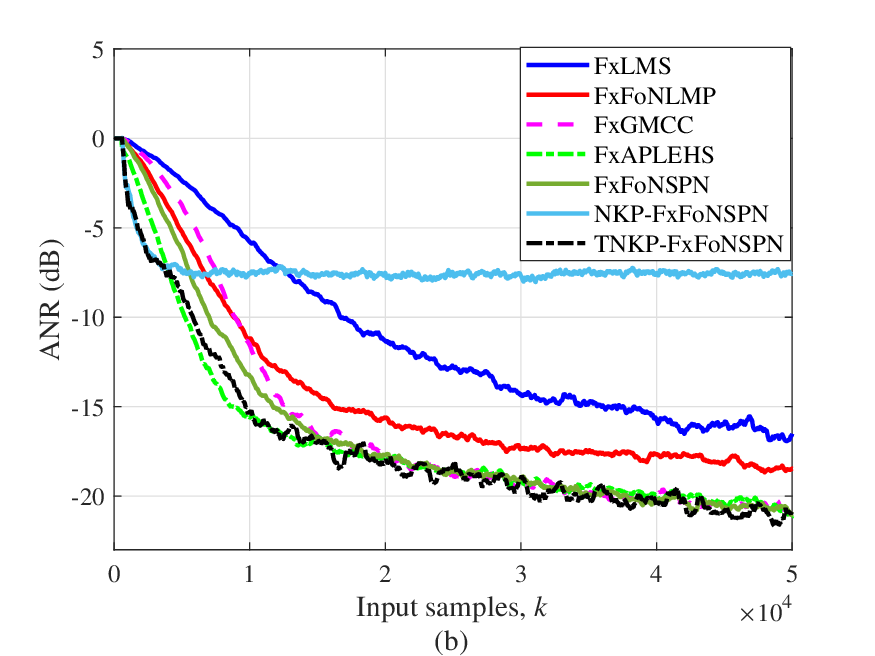}								  
	\caption{Comparison of ANC algorithms under pink noise. (a) The noise reduction results and (b) the ANR curves of several algorithms. }
	\label{Fig14}
\end{figure}
\begin{figure}
	\centering  
	\includegraphics[scale=0.42] {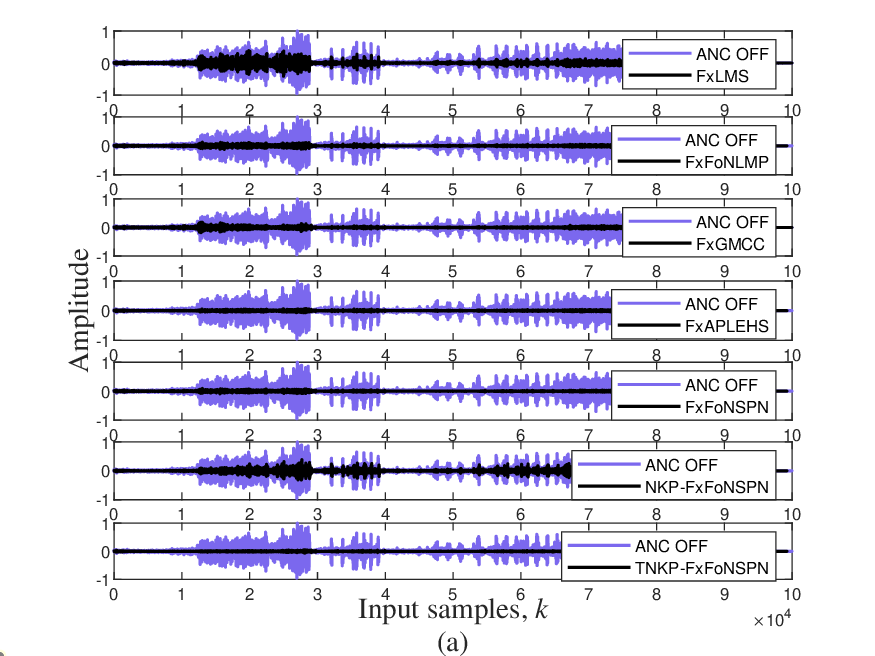}  
	\hspace{0.1ex}									 
	\includegraphics[scale=0.42] {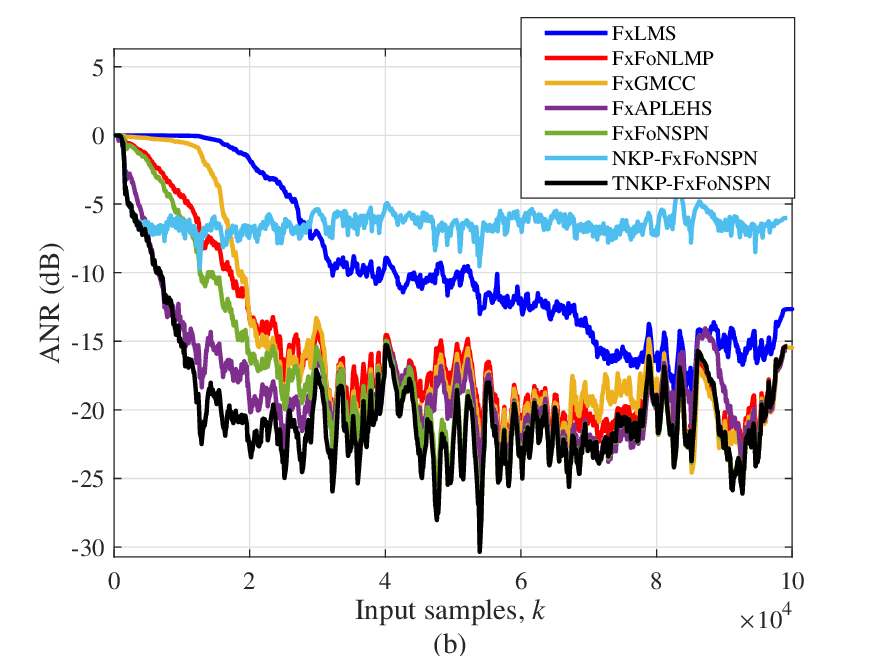}								  
	\caption{Comparison of ANC algorithms under real-world gunshots. (a) The noise reduction results and (b) the ANR curves of several algorithms. }
	\label{Fig15}
\end{figure}
\begin{table*}[htp]
	\renewcommand\arraystretch{1}
	\tabcolsep = 0.08cm
	
	\setlength{\abovecaptionskip}{0cm}
	\setlength{\belowcaptionskip}{0cm}
	\begin{center}
		\caption{Simulation parameters in active noise control.}
		\begin{tabular}{c c c c c} 
			\hline
			\rowcolor{gray!30}	Algorithms &Pink noise  &Gunshots &Helicopter noise & Pile driver noise  \\			
			\hline
			FxLMS	 &$\mu=0.01$	&$\mu=0.0004$&$\mu=0.01$ & $\mu=0.001$\\
			\rowcolor{gray!30}FxFoNLMP &$\mu=0.009$, $p=1.9$&$\mu=0.003$, $p=1.2$&$\mu=0.008$, $p=1.2$ & $\mu=0.002$, $p=1.2$\\
			\rowcolor{gray!30} & $\beta=0.9$&$\beta=1.1$&$\beta=1.1$ & $\beta=1.1$\\
			FxGMCC &$\mu=0.0007$, $p=1.1$ & $\mu=0.001$, $p=1.5$& $\mu=0.001$, $p=1.1$&  $\mu=0.001$, $p=1.1$\\
			&$\eta=0.2$ & $\eta=0.2$&$\eta=0.2$& $\eta=3$\\
			\rowcolor{gray!30}FxAPLEHS&$\mu=0.01$, $\chi=0.01$ & $\mu=0.01$, $\chi=0.01$ & $\mu=0.02$, $\chi=0.01$& $\mu=0.01$, $\chi=0.01$\\
			\rowcolor{gray!30}&$P=5$ & $P=8$ & $P=5$& $P=1$\\
			FxFoNSPN &$\mu=0.03$, $p=1.9$&$\mu=0.008$, $p=1.2$&$\mu=0.02$, $p=1.2$ & $\mu=0.005$, $p=1.2$\\
			& $\beta=0.9$&$\beta=1.1$&$\beta=1.1$ & $\beta=1.1$\\
			\rowcolor{gray!30}NKP-FxFoNSPN &$\mu=0.04$, $p=1.9$&$\mu=0.03$, $p=1.2$&$\mu=0.04$, $p=1.2$ &$\mu=0.01$, $p=1.2$\\
			\rowcolor{gray!30}&$Q=5$, $\beta=0.9$&$Q=5$, $\beta=1.1$&$Q=5$, $\beta=1.1$ &$Q=5$, $\beta=1.1$\\
			TNKP-FxFoNSPN &$\mu=0.04$, $p=1.9$&$\mu=0.03$, $p=1.2$&$\mu=0.04$, $p=1.2$ &$\mu=0.01$, $p=1.2$\\
			&$Q=5$, $\beta=0.9$, $\mu_b=0.03$&$Q=5$, $\beta=1.1$, $\mu_b=0.008$&$Q=5$, $\beta=1.1$, $\mu_b=0.002$ &$Q=5$, $\beta=1.1$, $\mu_b=0.005$\\
			\hline
		\end{tabular}
	\end{center}
\end{table*}

Fig. 17 shows the ANR learning curves and residual error signals for the FxLMS \cite{ elliott2000}, FxFoNLMP \cite{3}, FxGMCC \cite{zhu2020robust}, FxAPLEHS \cite{zhou2024combined}, FxFoNSPN, NKP-FxFoNSPN, and TNKP-FxFoNSPN algorithms under pink noise. As illustrated in Fig. 17 (b), the proposed NKP-FxFoNSPN algorithm achieves the fastest ANR convergence among the compared algorithms; however, it exhibits poor steady-state performance. Furthermore, the FxFoNLMP algorithm converges faster than the conventional FxLMS algorithm. Meanwhile, the FxGMCC algorithm attains a lower steady-state ANR than the FxFoNLMP algorithm. The FxFoNSPN algorithm converges faster than the FxGMCC algorithm, which is attributed to the SAF structure's inherent decorrelation capability for correlated signals. In addition, the FxAPLEHS algorithm benefits from optimal step-size control via the error signal, achieving both faster convergence and lower steady-state misadjustment. By comparison, the proposed TNKP-FxFoNSPN algorithm achieves comparable performance to that of the FxAPLEHS algorithm but demonstrates superior performance during the initial convergence stage. As depicted in Fig. 17 (a), the proposed TNKP-FxFoNSPN algorithm yields the best noise reduction results.

$B.$ {\it  Real-world Gunshots}
\begin{figure}
	\centering  
	\includegraphics[scale=0.42] {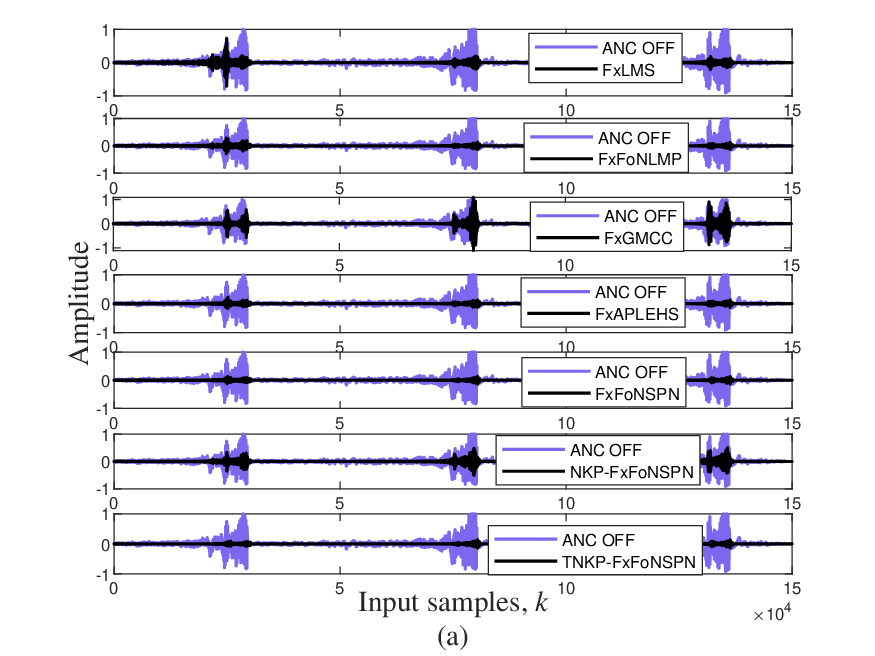}  
	\hspace{0.1ex}									 
	\includegraphics[scale=0.42] {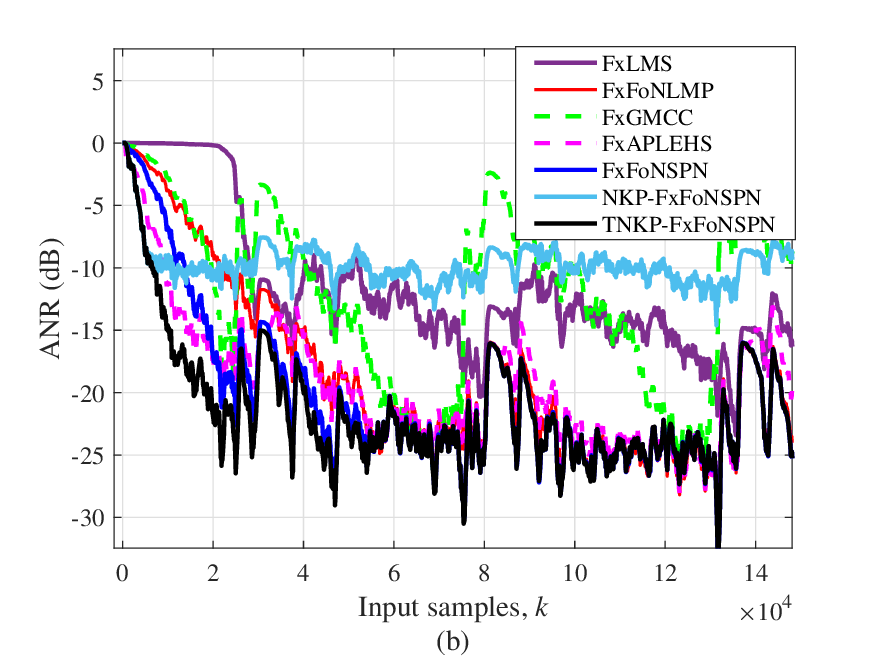}
	\hspace{0.1ex}	
	\includegraphics[scale=0.42] {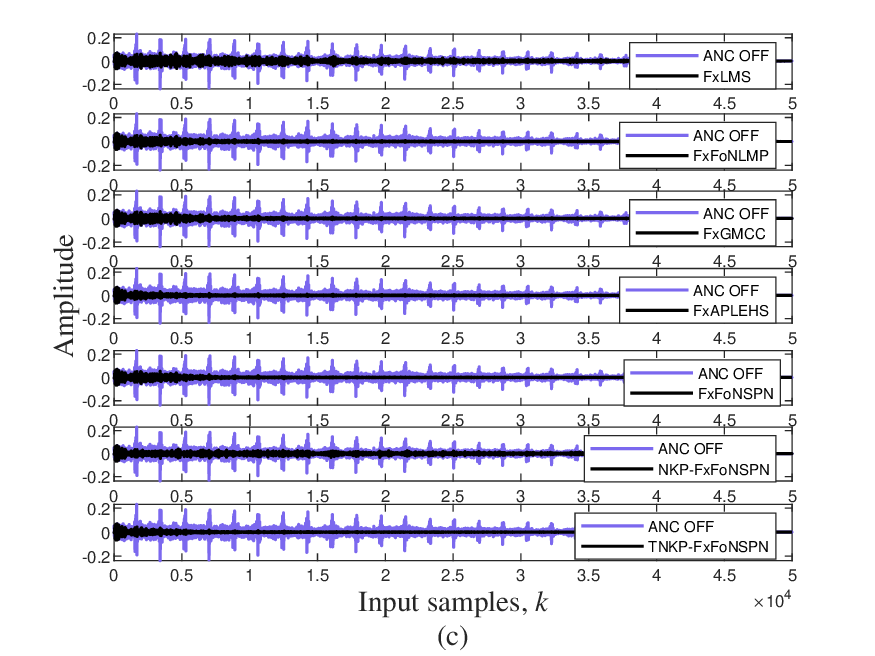}  
	\hspace{0.1ex}									 
	\includegraphics[scale=0.42] {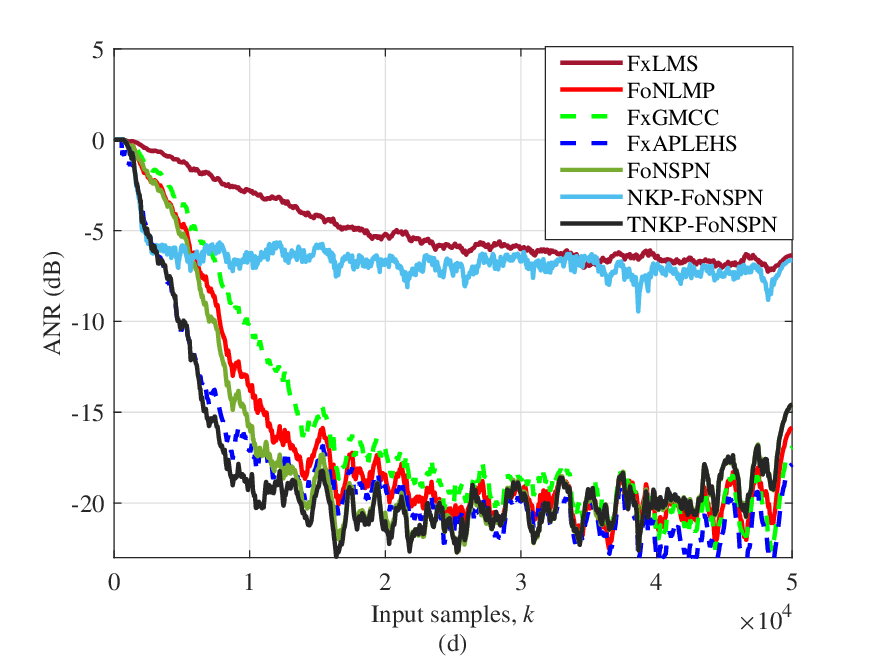}								  
	\caption{Performance comparison for different algorithms. (a) The noise reduction results of pile driver noise; (b) The ANR curves of several algorithms under pile driver noise; (c) The noise reduction results of helicopter noise; (d) The ANR curves of several algorithms under helicopter noise. }
	\label{Fig16}
\end{figure}
\begin{figure}
	\centering  
	\includegraphics[scale=0.42] {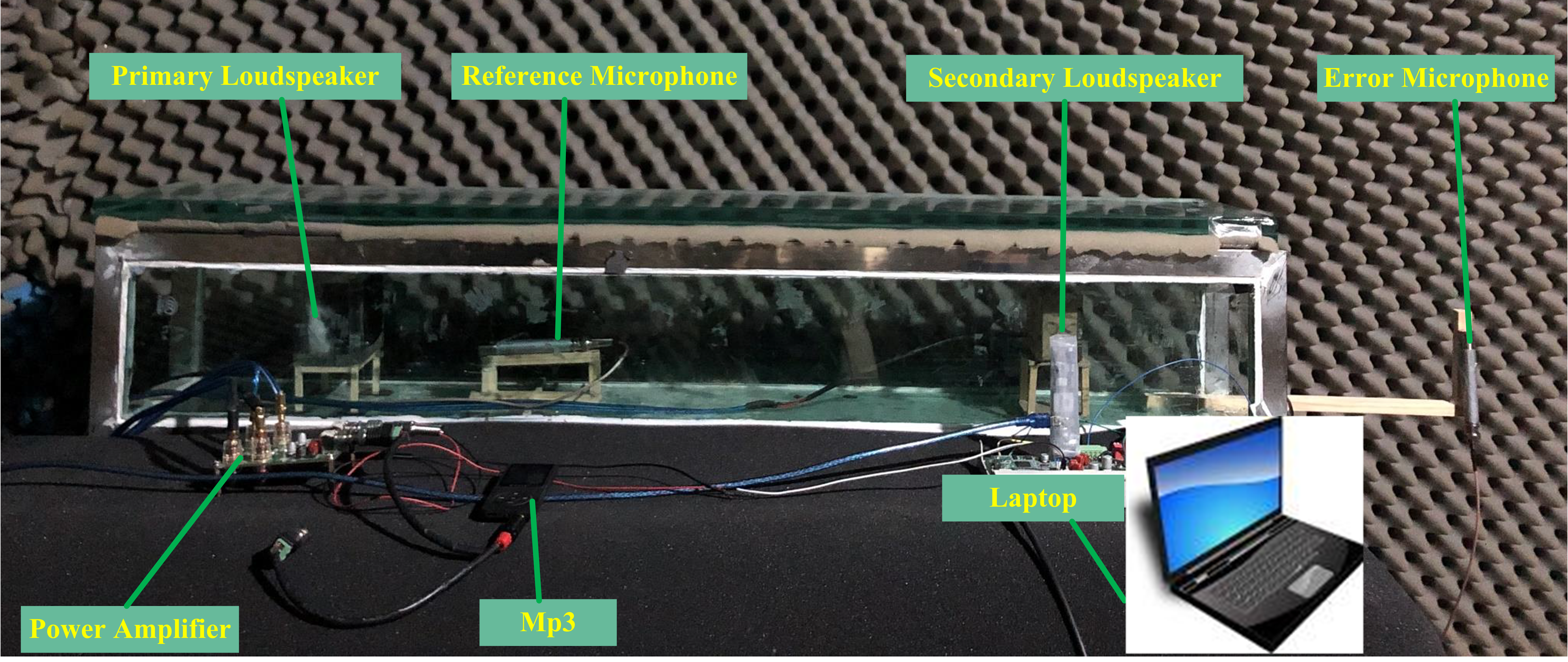}  								  
	\caption{The single-channel duct ANC system.}
	\label{Fig17}
\end{figure}
\begin{figure}
	\centering  
	\includegraphics[scale=0.42] {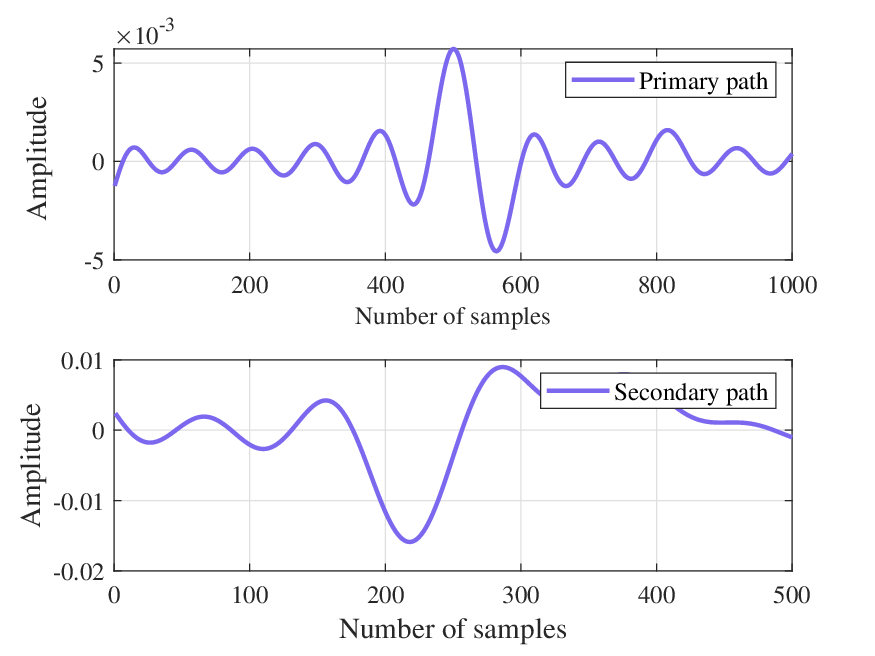}  								  
	\caption{Time-domain IRs of the simulated primary and secondary paths \cite{zhou2024combined}.}
	\label{Fig18}
\end{figure}
\begin{figure}
	\centering  
	\includegraphics[scale=0.42] {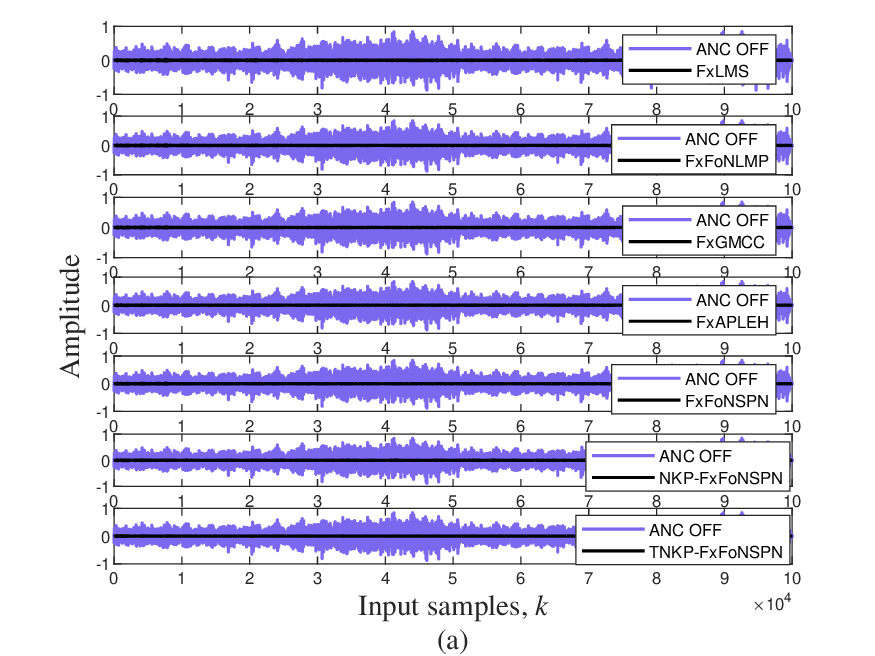}  
	\hspace{0.1ex}									 
	\includegraphics[scale=0.42] {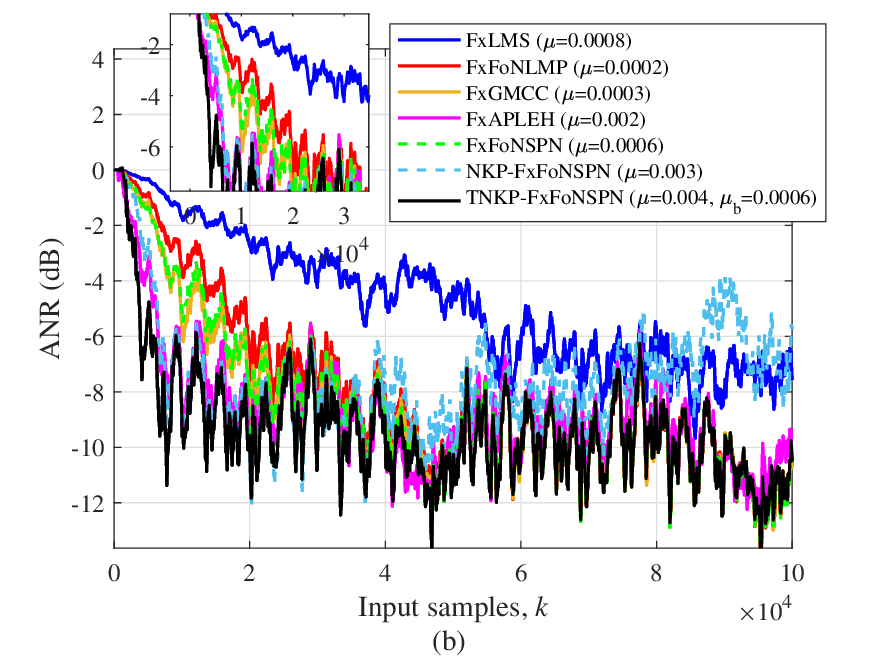}								  
	\caption{Comparison of ANC algorithms under real traction substation noise. (a) The noise reduction results and (b) the ANR curves of several algorithms. $p=0.9$ and $\beta=0.8$ for the FxFoMLNP algorithm; $p=1.5$ and $\eta=0.2$ for the FxGMCC algorithm; $\chi=0.01$ and $P=8$ for the FxAPLEHS algorithm; $p=0.9$ and $\beta=0.8$ for the FxFoNSPN algorithm; $Q=5$, $p=0.9$, and $\beta=0.8$ for the NKP-FxFoNSPN and TNKP-FxFoNSPN algorithms. }
	\label{Fig19}
\end{figure}

Fig. 18 presents the ANR learning curves and residual error signals for the algorithms under real-world gunshot noise, characterized by its non-Gaussian and impulsive nature. As shown in Fig. 18 (b), the FxFoNLMP and FxGMCC algorithms exhibit comparable overall ANR performance while converging faster than the conventional FxLMS algorithm. Furthermore, the FxFoNSPN algorithm leverages the inherent decorrelation properties of the SAF structure to achieve marginally faster convergence than the FxFoNLMP algorithm. Unfortunately, the proposed NKP-FxFoNSPN demonstrates rapid initial convergence but yields poor steady-state ANR performance. In contrast, both the FxAPLEHS and proposed TNKP-FxFoNSPN algorithms achieve superior ANR performance relative to other methods, while the latter converges more rapidly than the former. As shown in Fig. 18 (a), the proposed TNKP-FxFoNSPN algorithm achieves the best noise reduction results compared to the competing algorithms. 

$C.$ {\it  Real-world  helicopter noise and pile driver noise}

Fig. 19 further validates the effectiveness of the proposed algorithms using real-world helicopter and pile driver noise. Consistent with the resultings in Figs. 17 and 18, the proposed TNKP-FxFoNSPN algorithm demonstrates superior noise reduction performance compared to all competing methods.

$D.$ {\it  Real Duct ANC Acoustic Models}

We check the noise reduction performance of the proposed algorithms in a real single-channel duct ANC system. The noise source signal employed is a real noise recorded from a traction substation \cite{zhu2020robust}. The single-channel duct ANC system with a sampling frequency of 192kHz is placed in a semi-anechoic chamber, as shown in Fig. 20 \cite{10045592}. The system consists of a 3-inch primary loudspeaker (for playing the primary noise), two power amplifiers, a 1.8-inch secondary loudspeaker, and two types of 4966-H-041 pre-polarized 1/2-inch free-field microphones, which serve as reference and error sensors. These components are sensibly placed within a square section of duct 105 cm long and 14 cm wide. Furthermore, the time-domain IRs of the primary path with a length of 1000 and the secondary path with a length of 500 are shown in Fig. 21.

Fig. 22 examines the performance of the proposed algorithms and the compared algorithms in dealing with the real noise recorded at the traction substation. Consistent with previous simulations, the proposed TNKP-FxFoNSPN algorithm outperforms all competing methods in both convergence rate and steady-state misadjustment.

$E.$ {\it  Multi-channel ANC Model}
 \begin{figure}
 	\centering  
 	\includegraphics[scale=0.55] {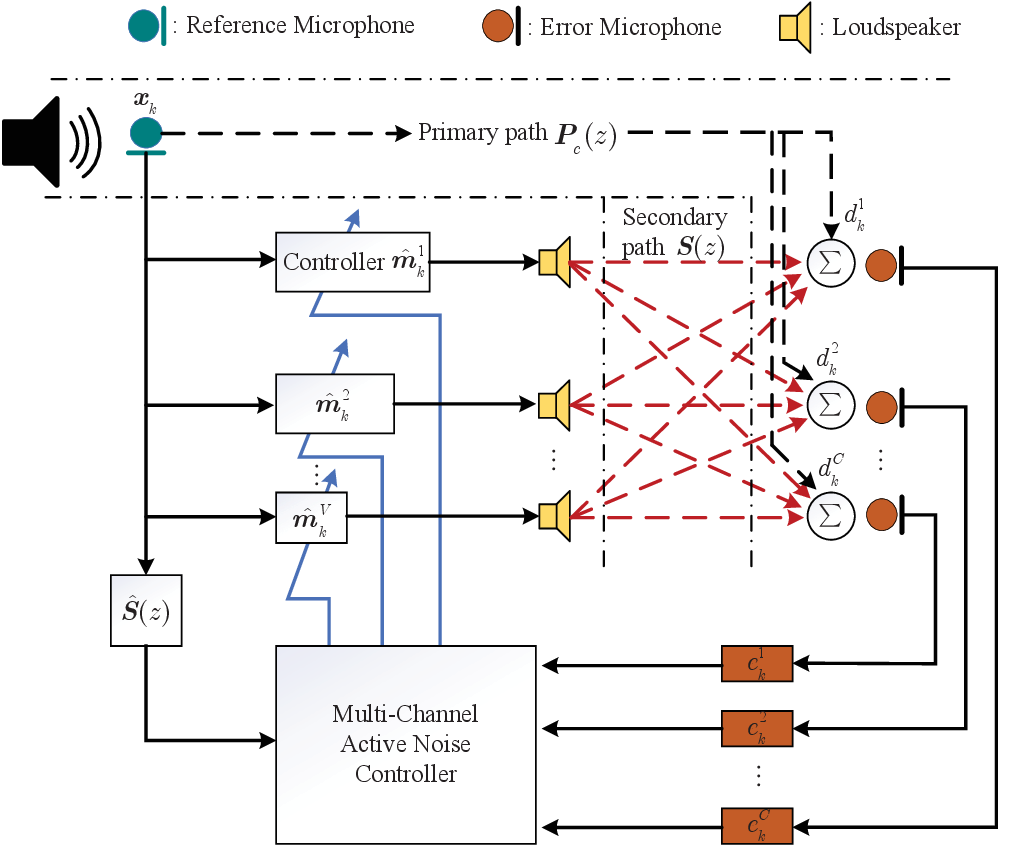}  								  
 	\caption{Diagram of a general multi-channel active noise control system.}
 	\label{Fig23}
 \end{figure}

In this subsection, we verified the reliability of the proposed algorithms in a multi-channel ANC system. Fig. 23 describes the diagram of a general multi-channel ANC system that includes the $V$ secondary loudspeaker, the $C$ error microphone, and a reference microphone. Unlike single-channel ANC systems, $\bm{P}_c(z)$ represents the primary acoustic path connecting the reference microphone to the $c$th error microphone. The impulse response of the secondary path connecting the $v$th secondary loudspeaker to the $c$th error microphone is $\bm{s}_{vc}$. Driven by noise source $\bm{x}_k$, the error signal $e_k^c$ measured by the $c$th error microphone can be expressed as 
\begin{equation}
	\begin{split}
		\begin{array}{rcl}
			\begin{aligned}
				\label{058}
				\ e_k^c=d_k^c-y_k^c=d_k^c-\sum_{v=1}^V \hat{\bm s}_{vc}*\big(\{\hat{\bm m}_k^v\}^{\text T}\bm{x}_k\big),
			\end{aligned}
		\end{array}
	\end{split}
\end{equation}  
where $d_k^c$ denotes the $c$th desired signal, $y_k^c$ is the output signal at the $c$th error microphone, $\hat{\bm s}_{vc}=[\hat{s}_{vc}^0,\hat{s}_{vc}^1,...,\hat{s}_{vc}^{D_s-1}]^{\text T}$ represents the estimated value of true impulse response $\bm{s}_{vc}$. Here, we difine $\bar{y}_k^c\overset{\bigtriangleup}{=}\{\hat{\bm m}_k^v\}^{\text T}\bm{x}_k$. Therefore, $y_k^c$ can be rewritten as  
\begin{equation}
	\begin{split}
		\begin{array}{rcl}
			\begin{aligned}
				\label{059}
				\ y_k^c&=\sum_{v=1}^V \hat{\bm s}_{vc}*\big(\{\hat{\bm m}_k^v\}^{\text T}\bm{x}_k\big)=\sum_{v=1}^V\sum_{d_s=0}^{D_s-1} \hat{s}_{vc}^{d_s}\bar{y}_{k-d_s}^c=\sum_{v=1}^V\sum_{d_s=0}^{D_s-1} \hat{s}_{vc}^{d_s}\{\hat{\bm m}_{k-d_s}^v\}^{\text T}\bm{x}_{k-d_s}\\&\approx \sum_{v=1}^V\sum_{d_s=0}^{D_s-1} \hat{s}_{vc}^{d_s}\{\hat{\bm m}_{k}^v\}^{\text T}\bm{x}_{k-d_s}=\sum_{v=1}^V \{\hat{\bm m}_{k}^v\}^{\text T}\bm{x}_k^v,
			\end{aligned}
		\end{array}
	\end{split}
\end{equation}
where the approximation operation is valid when the controller $\hat{\bm m}_k^v$ vary slowly, and $\bm{x}_k^v$ denotes the filtered input signal, which can be calculated by 
\begin{equation}
	\begin{split}
		\begin{array}{rcl}
			\begin{aligned}
				\label{060}
				\ \bm{x}_k^v=\sum_{d_s=0}^{D_s-1}\hat{s}_{vc}^{d_s}\bm{x}_{k-d_s}=\bm{X}_k\hat{\bm s}_{vc},
			\end{aligned}
		\end{array}
	\end{split}
\end{equation} 
where 
\begin{equation}
	\begin{split}
		\begin{array}{rcl}
			\begin{aligned}
				\label{061}
				\bm{X}_k=\big[\bm{x}_k,\bm{x}_{k-1},...,\bm{x}_{k-D_s+1}\big] \in \mathbb{R}^{D \times D_s}.
			\end{aligned}
		\end{array}
	\end{split}
\end{equation} 

In the multi-channel ANC scenario, we define $\hat{\bm m}_k\overset{\bigtriangleup}{=}\big[\{\hat{\bm m}_k^1\}^{\text T},\{\hat{\bm m}_k^2\}^{\text T},...,\{\hat{\bm m}_k^V\}^{\text T}\big]^{\text T}\in \mathbb{R}^{DV \times 1}$ and $\bar{\bm x}_k^v\overset{\bigtriangleup}{=}\big[\{\bm{x}_k^1\}^{\text T},\{\bm{x}_k^2\}^{\text T},...,\{\bm{x}_k^V\}^{\text T}\big]^{\text T}\in \mathbb{R}^{DV \times 1}$. Therefpre, \eqref{059} can be rewritten as
\begin{equation}
	\begin{split}
		\begin{array}{rcl}
			\begin{aligned}
				\label{062}
				y_k^c=\hat{m}_k^{\text T}\bar{\bm x}_k^v, \;v=1,2,...,V.
			\end{aligned}
		\end{array}
	\end{split}
\end{equation}

\textbf{Remark 7}. Since the multi-channel ANC system is a dimensional extension of the single-channel ANC system, and we have separately provided the pseudocodes of the NKP-FxFoNSPN and TNKP-FxFoNSPN algorithms in the single-channel ANC system in Tables 3 and 4, the update scheme of Controller $\hat{\bm m}_k^v$ in the multi-channel ANC system is similar to that in the single-channel scenario. Therefore, to avoid structural redundancy in the paper, we will not redundantly present the pseudocodes of the proposed algorithms in the multi-channel ANC scenario.

In multi-channel ANC scenarios, the ANR can be rewritten as \cite{chien2022affine} 
\begin{equation}
	\begin{split}
		\begin{array}{rcl}
			\begin{aligned}
				\label{063}
				\ \text{ANR}_s(k)\overset{\bigtriangleup}{=}20\text{log}_{10}\frac{1}{C}\sum_{c=1}^C\frac{Z_e^c(k)}{Z_d^c(k)},
			\end{aligned}
		\end{array}
	\end{split}
\end{equation}
where $ Z_e^c(k)=\eta Z_e^c(k-1) + (1-\eta)\lvert e_k^c\lvert$ and $Z_d^c(k)=\eta Z_d^c(k-1) + (1-\eta)\lvert d_k^c\lvert$ denote the averaged magnitude of residual error and desired signal at the $c$th error microphone. 
 \begin{figure}
	\centering  
	\includegraphics[scale=0.45] {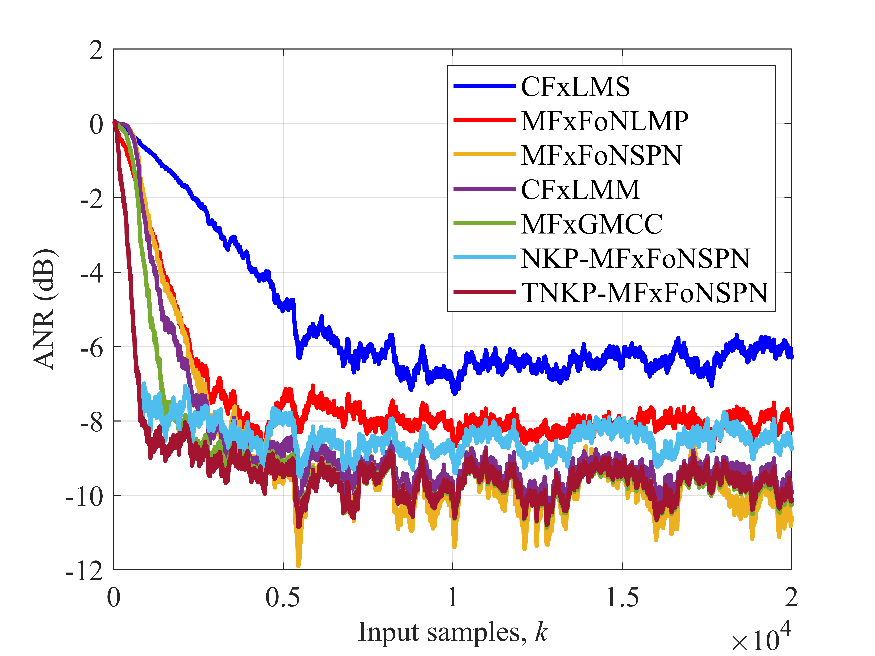}  								  
	\caption{The $\text{ANR}_s$ curves of the algorithms in the multi-channel active noise control system. $D=20$, $D_1=5$, $D_2=4$, and $Q=4$ for the NKP technology. $\mu=0.0009$ for the CFxLMS algorithm; $\mu=0.0004$, $p=1.1$, and $\eta=0.2$ for the MFxGMCC algorithm; $p=1.2$, $\beta=1.1$ and $\mu=0.001$ for the MFxFoNLMP algorithm; $p=1.2$, $\beta=1.1$, and $\mu=0.004$ for the MFxFoNSPN algorithm; $p=1.2$, $\beta=1.1$, and $\mu=0.009$ for the NKP-MFxFoNSPN algorithm; $\mu=0.009$, $\mu_b=0.004$, $p=1.2$, and $\beta=1.1$ for the TNKP-MFxFoNSPN algorithm.}
	\label{Fig24}
\end{figure}

Furthermore, the transfer functions of the secondary paths are $S_{11}(z)=z^{-3}-1.4z^{-4}$, $S_{21}(z)=z^{-3}-1.4z^{-4}$, $S_{31}(z)=z^{-3}-1.6z^{-4}$, $S_{41}(z)=z^{-3}-1.5z^{-4}$, $S_{51}(z)=z^{-3}-1.3z^{-4}$, $S_{12}(z)=z^{-3}-1.5z^{-4}$, $S_{22}(z)=z^{-3}-1.5z^{-4}$, $S_{32}(z)=z^{-3}-1.4z^{-4}$, $S_{42}(z)=z^{-3}-1.3z^{-4}$, $S_{52}(z)=z^{-3}-1.4z^{-4}$, $S_{13}(z)=z^{-3}-1.3z^{-4}$, $S_{23}(z)=z^{-3}-1.3z^{-4}$, $S_{33}(z)=z^{-3}-1.5z^{-4}$, $S_{43}(z)=z^{-3}-1.6z^{-4}$, $S_{53}(z)=z^{-3}-1.5z^{-4}$, $S_{14}(z)=z^{-3}-1.4z^{-4}$, $S_{24}(z)=z^{-3}-1.4z^{-4}$, $S_{34}(z)=z^{-3}-1.6z^{-4}$, $S_{44}(z)=z^{-3}-1.5z^{-4}$, $S_{54}(z)=z^{-3}-1.6z^{-4}$, $S_{15}(z)=z^{-3}-1.5z^{-4}$, $S_{25}(z)=z^{-3}-1.5z^{-4}$, $S_{35}(z)=z^{-3}-1.4z^{-4}$, $S_{45}(z)=z^{-3}-1.5z^{-4}$, and $S_{55}(z)=z^{-3}-1.4z^{-4}$. The transfer functions of the primary paths are $P_{1}(z)=1.5z^{-6}-0.2z^{-7}+0.3z^{-8}$, $P_{2}(z)=1.4z^{-6}-0.4z^{-7}+0.1z^{-8}$, $P_{3}(z)=1.6z^{-6}-0.3z^{-7}+0.2z^{-8}$, $P_{4}(z)=1.4z^{-6}-0.4z^{-7}+0.1z^{-8}$, and $P_{5}(z)=1.5z^{-6}-0.2z^{-7}+0.3z^{-8}$. In addition, the noise source signal employed is a real noise recorded from a traction substation \cite{zhu2020robust}.

As shown in Fig. 24, similar to the single-channel ANC scenario, the proposed multi-channel NKP-FxFoNSPN (NKP-MFxFoNSPN) algorithm converges faster than the centralized FxLMS (CFxLMS) \cite{morgan2003analysis} and multi-channel FxFoNLMP (MFxFoNLMP) algorithms. However, its steady-state error is higher than that of the multi-channel FxFoNSPN (MFxFoNSPN) algorithm. The convergence performance of the multi-channel FxGMCC (MFxGMCC) algorithm is superior to that of the centralized filtered-x least mean M-estimate (CFxLMM) \cite{zhou2025distributed} and MFxFoNSPN algorithms. In contrast, the proposed multi-channel TNKP-FxFoNSPN (TNKP-MFxFoNSPN) algorithm achieves a faster convergence speed than the NKP-MFxFoNSPN, CFxLMM, and MFxGMCC algorithms.
\section{Conclusion}
This paper proposes a novel NKP-FoNSPN algorithm for adaptive filtering in complex impulsive noise environments. Compared with the conventional NSPN, the proposed method exhibits superior robustness under non-Gaussian inputs and $\alpha$-stable noise ($0<\alpha\leq2$), while achieving enhanced convergence in sparse system identification. In addition, the NKP-FoNSPN algorithm is decomposed into two new variants: NKP-NSPN and FoNSPN. Furthermore, a TNKP decomposition technique with lower computational complexity than conventional NKP is designed for specific filter structures. Leveraging this advancement, the TNKP-FoNSPN algorithm achieves significantly reduced steady-state misadjustment and computational burden compared to the NKP-FoNSPN algorithm. For active noise control, filtered-x implementations (NKP-FxFoNSPN and TNKP-FxFoNSPN) are developed, and the derived NKP-FxNSPN and FxFoNSPN algorithms are established. Simulations using pink noise, helicopter noise, gunshots, and pile driver noise demonstrate the significant advantages of the proposed TNKP-FxFoNSPN algorithm. Experimental results from a real constructed single-channel duct ANC and a simulated multi-channel ANC systems further confirm their outperformance over state-of-the-art approaches.

\bibliographystyle{./elsarticle-num}
\bibliography{cas-refs}
\end{document}